\documentclass[aps,floatfix,prd,nofootinbib,superscriptaddress,reprint,showpacs,10pt,preprintnumbers,longbibliography]{revtex4-1}
\usepackage[utf8]{inputenc}
\usepackage[pdftex]{graphicx}
\usepackage{float}
\usepackage{amsmath}
\usepackage{amssymb}
\usepackage{mathtools}
\usepackage{booktabs}
\usepackage{siunitx}
\usepackage{amsfonts}
\usepackage{dsfont}
\usepackage{array}
\usepackage{bm}
\usepackage{mathrsfs}
\usepackage{pifont}
\usepackage{multirow}
\usepackage{upgreek}
\usepackage[dvipsnames]{xcolor}
\usepackage[pdftex,
  pdftitle={Put PDF title here},
  pdfauthor={Put ODF authors here},
  bookmarks,
  colorlinks,
  linkcolor=myblue,
  citecolor=mymagenta,
  menucolor=black,
  urlcolor=myblue,
  plainpages=false,
  pdfpagelabels,
  hypertexnames=false]{hyperref}
\usepackage{verbatim}
\usepackage{slashed}
\usepackage{cleveref}
\usepackage{ucs}
\usepackage{subfigure}
\usepackage{csquotes}

\newcommand{\pvec}{\mathbf{p}}
\newcommand{\MSbar}{\overline{\mathrm{MS}}}
\newcommand{\tins}{t_{\mathrm{ins}}}
\newcommand{\tsnk}{t_{\mathrm{s}}}
\newcommand{\nstout}{N_{\mathrm{stout}}}
\newcommand{\rstout}{\rho_{\mathrm{stout}}}

\definecolor{mymagenta}{RGB}{200, 0, 100}
\definecolor{myblue}{RGB}{45, 48, 146}

\usepackage{xargs}
\usepackage[colorinlistoftodos,prependcaption,textsize=footnotesize]{todonotes}
\newcommandx{\UW}[2][1=]{\todo[linecolor=blue,backgroundcolor=blue!25,bordercolor=blue,#1]{UW: #2}}

\graphicspath{{plots/}}

\begin{document}
\title{Quark and gluon momentum fractions in the pion and in the kaon}

\author{Constantia Alexandrou}
\affiliation{Department of Physics, University of Cyprus, Nicosia, Cyprus}
\affiliation{Computation-based Science and Technology Research Center, The Cyprus Institute, Nicosia, Cyprus}

\author{Simone Bacchio}
\affiliation{Computation-based Science and Technology Research Center, The Cyprus Institute, Nicosia, Cyprus}


\author{Martha Constantinou}
\affiliation{Department of Physics,  Temple University,  Philadelphia,  PA 19122 - 1801,  USA}

\author{Joseph Delmar}
\affiliation{Department of Physics,  Temple University,  Philadelphia,  PA 19122 - 1801,  USA}

\author{Jacob Finkenrath}
\affiliation{Department of Physics, Bergische Universit\"{a}t Wuppertal, Gaußstraße 20, Wuppertal, 42119, Germany}

\author{Bartosz Kostrzewa}
\affiliation{High Performance Computing and Analytics Lab, Rheinische Friedrich-Wilhelms-Universität Bonn,
Germany}


\author{Marcus Petschlies}
\affiliation{Helmholtz-Institut für Strahlen- und Kernphysik, University of Bonn, Germany}
\affiliation{Bethe Center for Theoretical Physics, University of Bonn, Germany}


\author{Luis Alberto Rodriguez Chacon}
\affiliation{Computation-based Science and Technology Research Center, The Cyprus Institute, Nicosia, Cyprus}
\affiliation{University of Ferrara, Ferrara, Italy}

\author{Gregoris Spanoudes}
\affiliation{Department of Physics, University of Cyprus, Nicosia, Cyprus}

\author{Fernanda Steffens}
\affiliation{Helmholtz-Institut für Strahlen- und Kernphysik, University of Bonn, Germany}
\affiliation{Bethe Center for Theoretical Physics, University of Bonn, Germany}

\author{Carsten Urbach}
\affiliation{Helmholtz-Institut für Strahlen- und Kernphysik, University of Bonn, Germany}
\affiliation{Bethe Center for Theoretical Physics, University of Bonn, Germany}

\author{Urs Wenger}
\affiliation{Institute for Theoretical Physics, Albert Einstein Center for Fundamental Physics, University of Bern, Switzerland}

\collaboration{Extended Twisted Mass Collaboration}

\date{\today}

\begin{abstract}
  We present the full decomposition of the momentum fraction carried by quarks and gluons in the  pion and the kaon.  We employ three gauge ensembles generated  with $N_f=2+1+1$ Wilson twisted-mass clover-improved fermions at the physical quark masses. For both mesons we perform a continuum extrapolation directly at the physical pion mass, which allows us to determine for the first time the momentum decomposition at the physical point. We find that the  total momentum fraction carried by quarks is 0.532(56) and 0.618(32) and by gluons 0.388(49) and 0.408(61) in the pion and in the kaon, respectively, in the $\MSbar$ scheme and at the renormalization scale of 2 GeV. 
  Having computed both the quark and gluon contributions in the continuum limit,  we find that the momentum sum is 0.926(68) for the pion and 1.046(90) for the kaon, verifying the momentum sum rule.
\end{abstract}

\maketitle

\textit{Introduction} ---
Quantum Chromodynamics (QCD) governs the formation 
of hadron bound states through the interaction of quarks 
and gluons. Understanding how these constituents 
combine to define the observable properties of 
hadrons has been the subject  
of extensive research spanning the last six decades. 
Among the diverse spectrum of hadrons, the light 
pseudo-scalar mesons hold a special place, as they 
encapsulate a key property of the QCD dynamics: 
they are recognized as pseudo-Goldstone 
bosons arising from the spontaneous breaking of  
chiral symmetry in QCD, characterized by their 
significantly smaller masses when compared to the 
more abundant nucleons. In addition, they stand
out for their valence quark-antiquark composition, 
in contrast to the structure of three valence quarks
in the nucleons. This inner structure is mostly
studied through the use of parton distribution
functions (PDFs), which measure, in the simplest case of 
unpolarized hadrons, the probability of
a parton to carry a fraction $x$ of the hadron momentum.
As a result, PDFs and their averages, the so-called
first moment of the PDFs,
are essential quantities to pin down the structure
of any hadron.

Among the pseudo-Goldstone 
bosons, pions and kaons are the simplest and the lightest. 
And while pions are comprised of valence up and 
down quarks and antiquarks, kaons are formed 
by combining either an up or a down quark (antiquark)
with a strange antiquark (quark). 
Thus, a significant
distinguishing feature between pions and kaons lies in the
mass of the strange quark, the mass of which exceeds the sum of 
the masses of the two light quarks by more than twenty fold,
while the pion mass is approximately one-fourth of the
kaon mass.
Moreover, in the
chiral limit their masses vanish, while the mass
of the nucleon is significantly less affected.
We may then say that comprehending 
the  pion and the kaon structures, their similarities
and differences, allows us to gain
a deeper understanding of QCD itself, 
possibly providing hints on the nonperturbative mechanisms 
behind the generation of hadron masses.
Indeed, the future Electron-Ion Collider 
(EIC)~\cite{Aguilar:2019teb,AbdulKhalek:2021gbh,Xie:2021ypc} 
has among its aims to increase
our knowledge of the partonic structure of pions
and kaons, and to quantify what are the quark and gluon
energy contributions to the mass of pions
and kaons.

Experimental data for the pion PDFs and their moments is scarce, 
restricted to limited data from several decades ago,
when they were measured
in pion-induced Drell-Yan processes~\cite{Conway:1989fs}.
For the kaon, the situation is even worse, with only
a handful of data~\cite{Saclay-CERN-CollegedeFrance-EcolePoly-Orsay:1980fhh}.
These early data has been included in modern global
analysis~\cite{Barry:2018ort,Novikov:2020snp,Barry:2021osv,JeffersonLabAngularMomentumJAM:2022aix,Kotz:2023pbu} and used to determine
the pion PDFs, while the kaon distributions have been studied
mostly within
models~\cite{Chen:2016sno,Shi:2018mcb,Lan:2019rba,Bednar:2018mtf,Cui:2020tdf,Han:2020vjp,Roberts:2021nhw,Pasquini:2023aaf}.
On the theory side, lattice QCD can investigate the
partonic structure of hadrons either through the 
direct computation of the PDFs~\cite{Constantinou:2020pek} or 
through their moments, employing the method of 
local operators. Based on this latter method, 
the body of lattice QCD studies on the 
pion structure remains relatively modest in comparison 
to those focused on the nucleon, a situation that is
more pronounced when considering studies on the structure 
of kaons. 
Many of the existing studies for the pion case 
either omit disconnected 
diagrams~\cite{Martinelli:1987zd,Best:1997qp,Guagnelli:2004ga,Capitani:2005jp,Abdel-Rehim:2015owa,Oehm:2018jvm,Alexandrou:2020gxs} or only 
incorporate a partial portion of the disconnected 
contributions~\cite{Loffler:2021afv}. 
The only exceptions are~\cite{ExtendedTwistedMass:2021rdx,Hackett:2023nkr}.
In our previous work~\cite{ExtendedTwistedMass:2021rdx}, 
we performed the first comprehensive breakdown 
of the pion momentum into its constituent quark and gluon 
components. The authors of~~\cite{Hackett:2023nkr} 
also presented a full momentum decomposition, albeit at larger than physical
pion mass and at a single lattice spacing.
While our previous work was conducted at the physical 
pion mass, it was also restricted to a singe lattice spacing.
In Ref.~\cite{Good:2023ecp}, 
an extrapolation to the continuum is made, but using ensembles
generated with heavier-than-physical pion masses. Moreover, their
result is restricted to the gluons only.
In the case of the kaon, there is one previous study 
\cite{Alexandrou:2020gxs}, where only connected contributions 
are considered and again a heavier than physical pion mass is used.
Going beyond the use of local 
operators, the last few years witnessed the development 
of a variety of new methods~\cite{Constantinou:2020pek}, 
based on nonlocal operators, 
which allow us to access the whole Bjorken $x$ dependence 
of PDFs. In particular, the $x$ dependence of PDFs for pions and kaons 
has been computed within the 
quasi-PDF~\cite{Zhang:2018nsy,Izubuchi:2019lyk,Lin:2020ssv,Gao:2020ito} 
and the pseudo-PDF~\cite{Joo:2019bzr,Salas-Chavira:2021wui} approaches,
as well as in the so called  \enquote{good lattice cross sections} method involving hadronic matrix elements computed on the lattice that are constructed to be time independent~\cite{Sufian:2019bol,Sufian:2020vzb}.
Nevertheless, these works are restricted to the valence quark
distributions, with the exception of~\cite{Salas-Chavira:2021wui} 
where only the gluon distribution is computed using ensembles 
with heavier than  physical pion masses. It is clear from this discussion that
the full momentum decomposition of the pion and of the kaon
is needed, both at the physical pion mass and with an 
extrapolation to the continuum. Therefore, in this work
we perform the first momentum
decomposition of both pions and kaons using a lattice
QCD computation at the physical point. \\

\textit{Lattice Setup} --- 
The parameters of the gauge field ensembles of the Extended Twisted Mass Collaboration (ETMC)
used in this work are listed 
in Table~\ref{tab:ensembles}, with the 
corresponding lattice spacing, lattice size
and two most relevant hadron masses
for charged pion and kaon.
All ensembles have $N_f=2+1+1$ quarks,
with masses of up, down, strange and charm quark 
tuned to reproduce the
physical pion mass, as well as the physical value of the ratios
$M_{D_s}/f_{D_s}$ and $m_c/m_s$, see Ref.~\cite{Alexandrou:2018egz} for details.
\begin{table}[th]
  \caption{ETMC ensembles analyzed. $a$ is the lattice spacing and $L\,(T=2L)$ the lattice spatial (temporal) extent in fm, and
    $M_{\pi^{\pm}}$ and $M_{K^{\pm}}$ the charged pion and kaon mass, respectively.
  }
\begin{tabular}{lccccc}
\hline
Ensemble & a [fm] & L~[fm] &  $M_{\pi^{\pm}}$~[MeV] & $M_{K^{\pm}}$~[MeV] \\ 
\hline
cB211.072.64 (B)&  $0.0796(1)$ & $5.09$  & $140.40(22)$  &  $498.41(11)$ \\
cC211.060.80 (C) &  $0.0682(1)$ & $5.46$  & $136.05(30)$   &  $495.27(14)$ \\
cD211.054.96 (D)&  $0.0569(1)$ & $5.46$  & $141.01(22)$  &  $494.77(11)$ \\
\hline
\end{tabular}
\label{tab:ensembles}
\end{table}
We extract the momentum fractions carried by 
the partons from the matrix elements of 
the energy-momentum tensor (EMT),
\begin{equation}
  \label{eq:x}
  \langle h(\pvec) \,|\, \bar{T}^X_{\mu\nu} \,|\, h(\pvec) \rangle =
  2\langle x\rangle_X\left(p_\mu p_\nu - \delta_{\mu\nu}\frac{p^2}{4}\right)
\end{equation}
where $h(\pvec)$ denotes a hadron state 
with mass $m_h$, and $p = (E_h(\pvec)\,,\,\,\pvec)$
the on-shell four-momentum, with
$E_h(\pvec) = \sqrt{m_h^2 + \pvec^2 }$.
The index $X = q, \, g$ denotes the
contributions from individual quark flavors ($q = u,\, d, \, s , \,c$) and gluons to the hadron momentum.
We compute the matrix
elements appearing on the left-hand side of Eq.~(\ref{eq:x})
through the ratio of three- and
two-point functions 
\begin{equation}
  \label{eq:Rdef}
  R_{\mu\nu}^X(\tins, \tsnk;\, \pvec)  \ =\ \frac{\langle h(\tsnk,\pvec)\,\bar{T}^X_{\mu\nu}(\tins)\,h(0,\pvec)\rangle}
  {\langle h(\tsnk,\pvec)\ h(0,\pvec)\rangle}.
\end{equation}
If $\tsnk$ and $\tsnk - \tins$ are large enough, then
asymptotically up to contamination by excited states
above the ground state $|h(\pvec)\rangle$
\begin{equation}
  \label{eq:RX}
  R_{\mu\nu}^X(t, t_f, t_i;\pvec) \to
  \frac{1}{2E_h}\, \frac{\langle h(\pvec)\, | \, \bar{T}^X_{\mu\nu} \, |\, h(\mathbf{p})\rangle}{1+\exp(-E_h(T
    - 2\,\tsnk ))}\,.
\end{equation}

The quark part of the EMT is given by 
\begin{equation}
  \bar{T}_{\mu\nu}^q\ =\ -\frac{(i)^{\kappa_{\mu\nu}}}{4}\,\bar q\,
  \left(
    \gamma_\mu\stackrel{\leftrightarrow}{D}_\nu + \gamma_\nu \stackrel{\leftrightarrow}{D}_\mu
    - \delta_{\mu\nu}\,\frac{1}{2}\,\gamma_\rho\,\stackrel{\leftrightarrow}{D}_\rho
  \right)q\,,
  \label{Tq}
\end{equation}
with $\kappa_{\mu\nu} = \delta_{\mu,4}\,\delta_{\nu,4}$, 
and $\stackrel{\leftrightarrow}{D}_\mu$ the symmetrized 
covariant derivative. The gluon EMT is given through the
gluon field-strength tensor $F_{\mu\nu}$ by
\begin{equation}
  \bar{T}^g_{\mu\nu} \ =\ (i)^{\kappa_{\mu\nu}}\,\left( 
  F_{\mu\rho}^{~} F_{\nu\rho} +
  F_{\nu\rho}^{~} F_{\mu\rho} - \delta_{\mu\nu}\,\frac{1}{2}\,F_{\rho\sigma}{~}F_{\rho\sigma}
  \right) \,.
  \label{Tg}
\end{equation}
We note that both expressions have been written for a Euclidean space-time.
While quark-connected and -disconnected diagrams contribute to the
3-point functions with quark EMT insertion, the gluonic EMT insertion produces only purely disconnected 
diagrams (cf. Fig.~\ref{fig:emt-3pt}
in the Supplementary Material \ref{sec:supplementaries}). \\

\begin{figure}
    \centering
    \includegraphics[width = 0.45\textwidth]{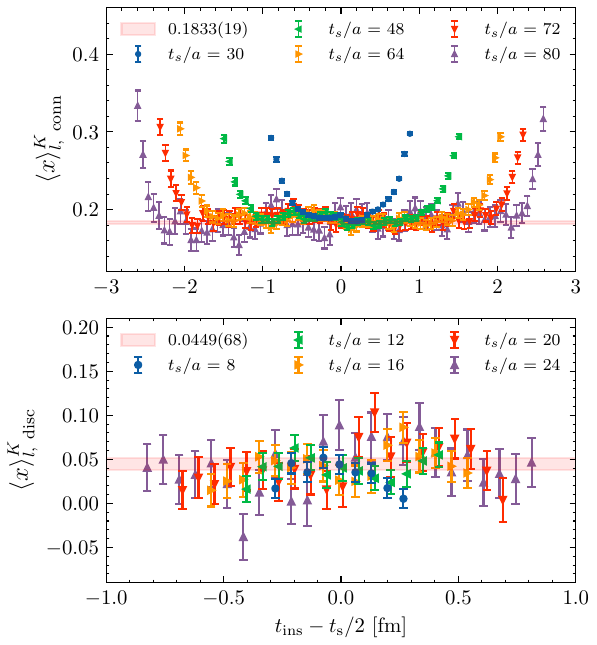}
    \caption{Bare ratios of the light-quark momentum fraction in the kaon dependence on the sink-source time separation  $\tsnk$ for the C ensemble. In the upper panel, we show the ratio for the connected contribution and in the lower panel, the ratio for the disconnected contribution.  We show results for several values of $\tsnk$, from $\tsnk/a=30$ to $\tsnk/a=80$. The red bands show the result after model averaging of fits  to a constant when varying the ranges of $\tsnk$ used. 
    }
    \label{fig:ratio_quark}
\end{figure}

\begin{figure}
   \centering
   \includegraphics[width = 0.45\textwidth]{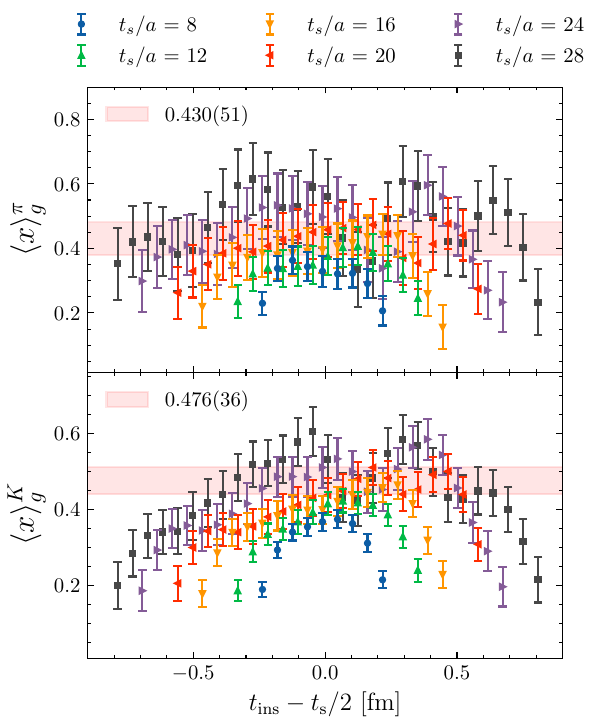}
   \caption{Examples of the bare ratios for the gluon contribution, for six values of the time sink-source  separation $t_s$ using 10 stout-smearing steps. 
   The top panel shows the pion and the bottom panel the kaon case. The horizontal bands
   show the model-averaged fit results.}
\label{fig:bare_ratio}
\end{figure}

\textit{Bare momentum fractions} ---
Equation (\ref{eq:RX}), combined with Eq.~(\ref{eq:x}),
allows us to determine $\langle x \rangle$ from 
different components of the EMT. We compute the 
connected contributions using the tensor element $\bar T_{44}$
with hadron momentum $\pvec = 0$,
while for the disconnected contributions we employ
the combination of components $\bar T_{4k}$, $k = 1,2,3$,
with nonzero momentum $|\pvec| = 2\pi/L$.
We thereby optimize the 
signal-to-noise ratio for each type of contribution:
the temporal-spatial EMT elements $\bar T_{4k}$
do not require additive renormalization by trace subtraction and with 
minimal nonzero hadron momentum allow for
most precise projection on the hadron ground state, necessary
for the weak signal in the disconnected diagrams.
On the other hand, statistical noise from trace subtraction is
negligible for the connected diagrams from $\bar T_{44}$, while 
zero hadron momentum renders the ground state
projection most precise. By Eq.~(\ref{eq:x}) 
and Lorentz-covariance, the matrix elements
from different tensor components become
equivalent in the continuum limit.
As an illustration of the data, we show the ratios 
for connected and disconnected 
contributions for the light quarks in the kaon 
in Fig.~\ref{fig:ratio_quark} for the 
ensemble C in Table~\ref{tab:ensembles}.
The ratios were computed for several source-sink separations $\tsnk$, and we see in Fig.~\ref{fig:ratio_quark} 
the formation of a plateau for large values of $\tsnk$.
We fitted the plateau using different combinations of 
$\tsnk$ and of the insertion time $\tins$. 
Model averaging~\cite{Jay:2020jkz} is used on the results from the 
plateau fit in order to obtain $\langle x \rangle$.
The same procedure is employed 
for the ratios involving the gluon EMT insertion.
To optimize the signal-to-noise ratio
we apply four-dimensional stout smearing \cite{Morningstar:2003gk} to the
gauge field in the lattice gluon field-strength tensor
when constructing $\bar T^g_{4k}$.
In Fig.~\ref{fig:bare_ratio} we show the ratio for 
several values of $\tsnk$ for $\nstout = 10$
smearing steps,
for both pion and kaon case.

We probe the systematic dependence on the
level of stout smearing by comparing the 
average momentum fraction for the range
of stout-smearing steps 
$5 \le \nstout \le 10$, weighted by the
renormalization factor $Z_{gg}$, which
is determined for the corresponding 
values of $\nstout$.
As we find agreement 
among those individual results, we then perform a simultaneous 
fit to the $Z_{gg}$-weighted ratios from this set of stout-smearing steps.
All the plateau fits, as well as the results 
after model averaging, for each ensemble and flavor, 
are shown in the Supplemental Material \ref{sec:supplementaries}. 
\\

\textit{Renormalization and continuum limit} ---
We proceed with the renormalization procedure 
for the flavor-singlet and flavor-nonsinglet combinations. 
The renormalized results are given in the  $\MSbar$ scheme
at scale 2 GeV.
The flavor-nonsinglet combinations are renormalized 
with the nonperturbatively determined 
renormalization factor $Z_{qq}$ as 
\begin{equation}
  \label{eq:nonsinglet}
  \begin{split}
    \langle x\rangle_{u+d-2s,\mathrm{R}} &= Z_{qq}(\langle x \rangle_u + \langle x
    \rangle_d - 2\langle x\rangle_s)\,,\\
    \langle x\rangle_{u+d+s-3c, \mathrm{R}} &= Z_{qq}(\langle x \rangle_u + \langle x
    \rangle_d + \langle x\rangle_s - 3 \langle x \rangle_c)\,.\\
  \end{split}
\end{equation}
Compared to our previous analysis~\cite{ExtendedTwistedMass:2021rdx}, 
we now use $\mathcal{O}(\alpha_s^4)$
corrections to the anomalous dimension, thus improving the scale
evolution of the data to the common $2$ GeV scale used throughout this
work.
The quark-singlet and gluon components mix under renormalization
according to 
\begin{equation}
  \label{eq:mixing}
  \begin{pmatrix}
    \langle x\rangle_{q,\mathrm{R}}\\
    \langle x\rangle_{g,\mathrm{R}}
  \end{pmatrix}
  =
  \begin{pmatrix}
    Z_{qq}^s & Z_{qg}\\
    Z_{gq} & Z_{gg} \\
  \end{pmatrix}
  \begin{pmatrix}
    \langle x\rangle_q\\
    \langle x\rangle_g
  \end{pmatrix}
\end{equation}
with $Z_{qq}^s$ the quark-singlet renormalization constant
and the total bare (renormalized) quark contribution is denoted by
\begin{align}
  \langle x\rangle_{q,(\mathrm{R})} &= \sum\limits_{f =u,d,s,c}\, \langle x\rangle_{f,(\mathrm{R})}\,.
  \label{eq:x-quark-total}
\end{align}
Defining $\delta Z_{qq} = Z_{qq}^s - Z_{qq}$,  one can solve 
Eq.~(\ref{eq:mixing}) for each single flavor and the gluon
component: 
\begin{equation}
  \label{eq:separate}
  \begin{split}
    \langle x\rangle_{f,\mathrm{R}} &= Z_{qq}\, \langle x\rangle_f +
    \frac{\delta Z_{qq}}{N_f}\, \sum_{f'} \, \langle x\rangle_{f'} +
    \frac{Z_{qg}}{N_f}\, \langle x \rangle_g\,,\\
    \langle x\rangle_{g,\mathrm{R}} &= Z_{gg}\, \langle x\rangle_g + Z_{gq}\,\sum_{f'} \langle x\rangle_{f'}\,.\\
  \end{split}
\end{equation}
The calculation of $Z_{gq}$ and $Z_{qg}$ have 
also been improved, with the details
presented in the Supplemental Material.

\begin{figure}
   \centering
   \includegraphics[width = 0.5\textwidth]{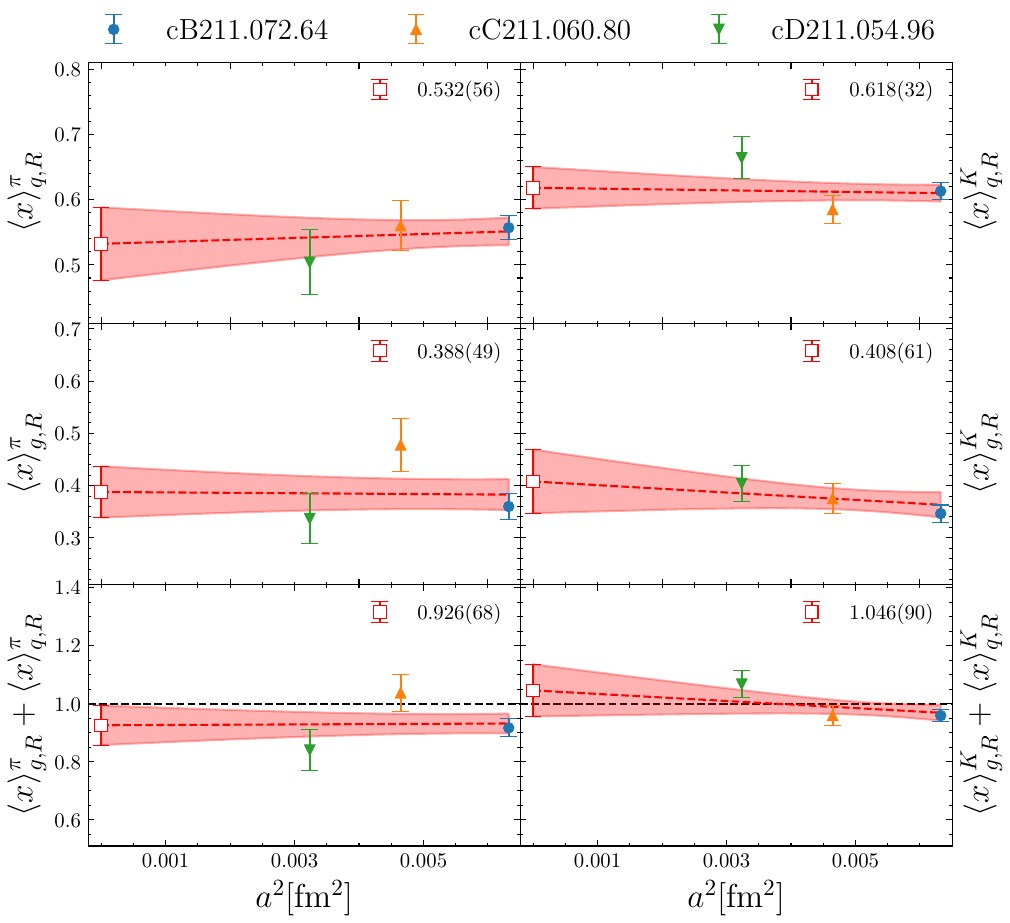}
   \caption{Continnum limit extrapolation for the pion (left panel) and the kaon (right panel). We present our results for the total quark and gluon contributions, as well as the momentum sum rule. The blue filled circles are results for ensemble B~\cite{ExtendedTwistedMass:2021rdx}, the orange upwards triangles for ensemble C and the green downwards triangles for ensemble D. The open symbol is the result after model averaging between constant and linear fit. Further details can be found in the Supplemental Material.
   }
   \label{fig:cl}
\end{figure}

For the twisted-mass lattice action at maximal twist
with automatic $\mathcal{O}(a)$ improvement
we expect leading discretization artefacts at second order
in the lattice spacing $a$.
We, thus, take the continuum limit using a constant or a linear fit in 
$a^2$, for both the
renormalized quark and gluon contributions, 
for the pion and the kaon cases. Our final value is the result of the model average among these fits.
In Fig.~(\ref{fig:cl}), we show the linear extrapolation 
of the total renormalized quark contribution 
$\langle x \rangle_{q, \mathrm{R}}$, 
the gluon contribution $\langle x \rangle_{g, \mathrm{R}}$, and 
in addition
their sum. We note that the total quark 
contribution to $\langle x \rangle^K$ tends
to be higher than its analog in the pion, as also
observed in Ref~\cite{Bednar:2018mtf}.
The gluon contribution to the
pion and to the kaon momenta, however, are compatible within
errors, and the momentum sum rule 
\begin{align}
  \langle x \rangle_{q, \mathrm{R}} + \langle x \rangle_{g, \mathrm{R}} = 1
  \label{eq:sum-rule}
\end{align}
is satisfied, in both cases, within a standard deviation.
The full list of the renormalized results in the continuum 
limit, for both pion and kaon, is presented in 
Table~\ref{tab:results_ours} at the scale of $\mu = 2$~GeV
in the $\MSbar$ scheme. 

\begin{figure}
    \centering
    \includegraphics[width = 0.49\textwidth]{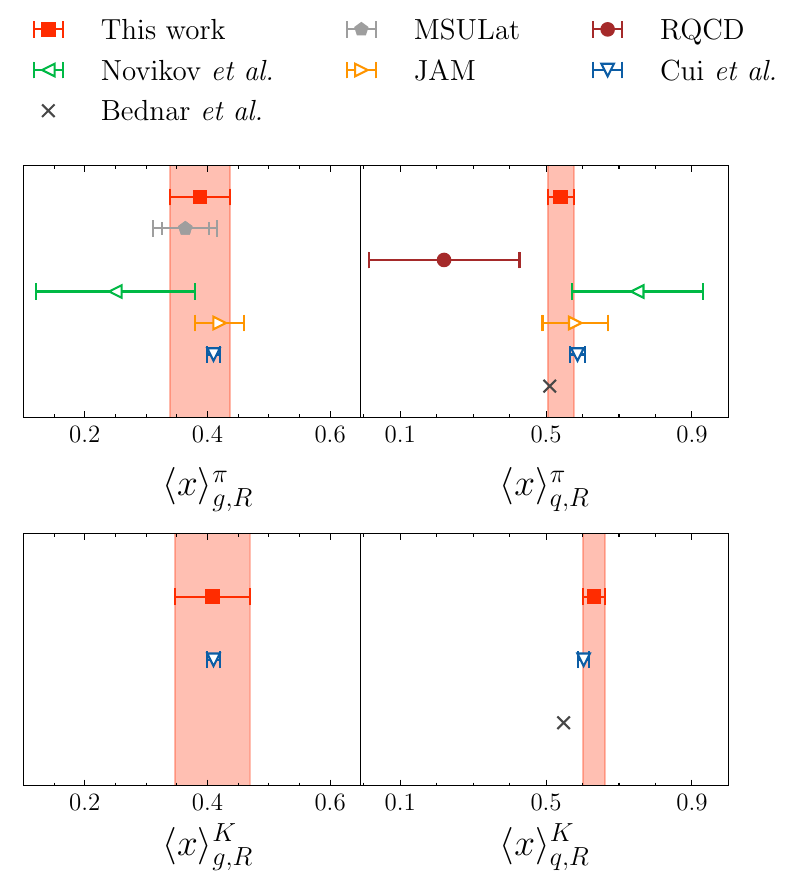}
    \caption{Comparison of the results of this work, with other available data, both from phenomenology and from lattice QCD.  All results are given in the $\MSbar$ scheme at the scale of $\mu=2\ \mathrm{GeV}$. The upper panels show the results for the pion gluon (left) and quark (right) momentum fractions, $\langle x \rangle_\textrm{g,R}^\pi$ and  $\langle x \rangle_\textrm{q,R}^\pi$, respectively. The lower panels show the corresponding results for the kaon.  The red filled squares show the results of this work with the red band the associated error band.  Recent results from phenomenological analyses of PDFs data are given by open symbols: left green triangle  (Novikov {\it et al.}~\cite{Novikov:2020snp}) and  right orange triangle (JAM Collaboration~\cite{Barry:2021osv}). The result  while the result based on the LFWF is represented by the down blue triangle (Cui {\it et al.}~\cite{Cui:2020tdf}), while using the DSE~\cite{Bednar:2018mtf}  by the black cross, where no error is provided. Recent lattice QCD results extrapolated to the continuum limit are given by  the brown filled circle (RQCD~\cite{Loffler:2021afv}) and the gray pentagon (MSULat~\cite{Good:2023ecp}).}
    \label{fig:final_comparison}
\end{figure}

In Fig.~\ref{fig:final_comparison}, we show 
a comparison of our results with those from other 
groups. We include recent results
from  phenomenological analyses~\cite{Novikov:2020snp,Barry:2021osv},  
from the Dyson Schwinger equations (DSE)~\cite{Bednar:2018mtf},
from a calculation using the light-front wave function (LFWF) approach~\cite{Cui:2020tdf} 
and from lattice QCD computations~\cite{Loffler:2021afv,Good:2023ecp},
limiting ourselves to those which are extrapolated to the 
physical pion mass and/or the continuum limit. 
The lattice QCD results shown are computed using 
ensembles simulated with pion masses larger than physical 
and then extrapolated to the physical pion mass. Furthermore,   
they both  considered only a partial sum of disconnected contributions, 
namely only the quark disconnected is considered in Ref.~\cite{Loffler:2021afv} 
but not the  gluon contribution, whereas only the gluon is considered  in Ref.~\cite{Good:2023ecp}.
For the cases where all contributions are included, the data is restricted
to only one lattice spacing~\cite{ExtendedTwistedMass:2021rdx} 
at physical pion mass or using larger than physical 
pion masses~\cite{Hackett:2023nkr, Alexandrou:2020gxs}. 
Our result for $\langle x \rangle_{g,R}^{\pi}$ is in agreement 
with the one from Ref.~\cite{Good:2023ecp} as is our result for  $\langle x \rangle_{q,R}^{\pi}$ with RQCD~\cite{Loffler:2021afv} that carries however a large error. 
Compared to our
previous work~\cite{ExtendedTwistedMass:2021rdx} using only the B ensemble, taking the continuum extrapolation increases the error, as can be seen in Fig.~\ref{fig:cl}.
For the kaon, the only available lattice QCD data~\cite{Alexandrou:2020gxs}
is for one lattice spacing using an ensemble with a heavier-than-physical 
pion mass and thus one cannot directly compare to 
the present work. \\

\begin{table}[th]
  \caption{Compilation of results for the pion and for the kaon in the continuum limit. All quantities are presented at the scale 
  $2\ \mathrm{GeV}$ in the $\MSbar$ scheme.}
  \centering
  \begin{tabular*}{.49\textwidth}{@{\extracolsep{\fill}}llr}
    \toprule\hline
    & $\pi$ & K \\
    \midrule\hline
    \(\langle x\rangle_{l,\mathrm{R}}\) & $ 0.448(34) $ & $0.260(09)$ \\
    \(\langle x\rangle_{s,\mathrm{R}}\) & $ 0.043(15) $ & $0.333(11)$ \\
    \(\langle x\rangle_{c,\mathrm{R}}\) & $ 0.019(17) $ & $0.024(17)$ \\
    \(\langle x\rangle_{g,\mathrm{R}}\) & $ 0.388(49) $ & $0.408(61)$  \\
    \(\langle x\rangle_{q,\mathrm{R}}\) & $ 0.532(56) $ & $0.618(32)$ \\
    \(\langle x \rangle_{u+d-2s,\mathrm{R}}\) & $ 0.382(17) $ & $-0.409(16)$ \\
    \(\langle x \rangle_{u+d+s-3c,\mathrm{R}}\) & $ 0.445(48) $ & $0.487(39)$ \\
    \bottomrule\hline
  \end{tabular*}
  \label{tab:results_ours}
\end{table}

%

\textit{Conclusions and outlook} ---
We present in this work the first complete momentum decomposition
for both the pion and the kaon in terms of their quark
and gluon consituents, performed within lattice QCD at 
the physical point. 
Three ensembles with $N_f=2+1+1$ quark flavors with
their masses tuned to reproduce the physical light, 
strange and charm quark masses are analyzed.  
We use a model average for the continuum extrapolation 
performed using both a constant and a linear fit
in the squared lattice spacing, $a^2$. 
Our results for $\langle x \rangle_g^{\pi,K}$ 
indicate a similar momentum fraction carried by gluons 
in the kaon and the pion. However, they
have  larger errors than the corresponding quark momentum fraction 
$\langle x \rangle_q^{\pi,K}$, which tends to be smaller 
in the pion. This indicates a possible larger momentum 
fraction carried by gluons in the pion as compared to the kaon.
Thus, among possible improvements, in future work  both  the statistical and systematic 
errors, in particular for $\langle x \rangle_g$, will be reduced. 
This can be accomplished by
increasing  the number of gauge configurations and source positions 
used for the two-point functions, Eq.~(\ref{eq:RX}), improving the statistical accuracy in the determination of the 
gluon momentum fraction. In addition, we plan to analyze 
an additional ensemble closer to the continuum limit in order to 
help a better error control when performing the continuum extrapolation. 
However, the current work presents  a remarkable achievement 
as we now have a fully consistent theoretical calculation
that allows us to directly compare the pion and the kaon
at the level of their quark and gluon structure. 
In particular, the momentum 
sum rule can be tested by computing all components from first principles. 
We stress that prior to the current work, 
there was no such decomposition into quark and gluon parts 
available of $\langle x \rangle^K$ using
a first-principles calculation.
Furthermore, the current work demonstrates clearly that the momentum 
fraction carried by gluons and sea quarks (the disconnected contributions) 
are important components of
the two lightest pseudo-Goldstone bosons, and added together with the 
valence contributions verify the momentum sum rule for both cases.

\begin{acknowledgments}
    G.S. acknowledges financial support from the European Regional Development Fund and the Republic of Cyprus through the Cyprus Research and Innovation Foundation under contract number EXCELLENCE/0421/0195.
    This project is partly funded by the European Union’s Horizon 2020 Research and Innovation Programme ENGAGE under the Marie Sklodowska-Curie COFUND scheme with grant agreement No. 101034267. C.A. acknowledges partial support from the European Regional Development Fund and the Republic of Cyprus through the Cyprus Research and Innovation Foundation under contract number EXCELLENCE/0421/0043 and the European Joint Doctorate project AQTIVATE funded by the European Commission under the Marie Sklodowska-Curie Doctoral Networks action and Grant Agreement No 101072344. This work is supported by the Swiss National Science Foundation (SNSF) through grant No.~200021\_175761, 200021\_208222, and 200020\_200424,
    as well as by the DFG and the NFSC as part of the Sino-German Collaborative Research Center CRC110 \enquote{Symmetries and the emergence of structure}. J. D. and M.~C. acknowledge financial support from the U.S. Department of Energy, Office of Nuclear Physics, Early Career Award under Grant No.\ DE-SC0020405.
    We gratefully acknowledge computing time granted on Piz Daint at Centro Svizzero di Calcolo Scientifico (CSCS)
    via the projects s849, s982, s1045, s1133 and s1197 and 
    the Gauss Centre for Supercomputing e.V. (www.gauss-centre.eu) for funding 
    this project by providing computing time on the GCS Supercomputers SuperMUC-NG 
    at Leibniz Supercomputing Centre and JUWELS \cite{JUWELS} at Juelich Supercomputing 
    Centre. The authors acknowledge the Texas Advanced Computing Center (TACC) at 
    The University of Texas at Austin for providing HPC resources (Project ID PHY21001). J.F. is
    supported by the DFG research unit FOR5269 ”Future methods for studying confined gluons 
    in QCD".
    The authors gratefully acknowledge PRACE for awarding access to HAWK at HLRS within the project with Id Acid 4886, and the Swiss National Supercomputing Centre (CSCS) and the EuroHPC Joint Undertaking for awarding this project access to the LUMI supercomputer, owned by the EuroHPC Joint Undertaking, hosted by CSC (Finland) and the LUMI consortium through the Chronos programme under project IDs CH17-CSCS-CYP and CH21-CSCS-UNIBE as well as the EuroHPC Regular Access Mode under project ID EHPC-REG-2021R0095.
    The open source software packages 
    tmLQCD~\cite{Jansen:2009xp,Abdel-Rehim:2013wba,Deuzeman:2013xaa}, 
    Lemon~\cite{Deuzeman:2011wz}, 
    QUDA~\cite{Clark:2009wm,Babich:2011np,Clark:2016rdz}, R~\cite{R:2019},
    cvc \cite{CVC:2024} and plegma have been used. Finally, 
    We thank the authors of~\cite{Cui:2020tdf} and~\cite{Bednar:2018mtf} for kindly sending us 
    their results evolved to the scale used in our 
    computation, $2\ \mathrm{GeV}$.

\end{acknowledgments}

\bibliography{bibliography}

\begin{thebibliography}{55}%
\makeatletter
\providecommand \@ifxundefined [1]{%
 \@ifx{#1\undefined}
}%
\providecommand \@ifnum [1]{%
 \ifnum #1\expandafter \@firstoftwo
 \else \expandafter \@secondoftwo
 \fi
}%
\providecommand \@ifx [1]{%
 \ifx #1\expandafter \@firstoftwo
 \else \expandafter \@secondoftwo
 \fi
}%
\providecommand \natexlab [1]{#1}%
\providecommand \enquote  [1]{``#1''}%
\providecommand \bibnamefont  [1]{#1}%
\providecommand \bibfnamefont [1]{#1}%
\providecommand \citenamefont [1]{#1}%
\providecommand \href@noop [0]{\@secondoftwo}%
\providecommand \href [0]{\begingroup \@sanitize@url \@href}%
\providecommand \@href[1]{\@@startlink{#1}\@@href}%
\providecommand \@@href[1]{\endgroup#1\@@endlink}%
\providecommand \@sanitize@url [0]{\catcode `\\12\catcode `\$12\catcode `\&12\catcode `\#12\catcode `\^12\catcode `\_12\catcode `\%12\relax}%
\providecommand \@@startlink[1]{}%
\providecommand \@@endlink[0]{}%
\providecommand \url  [0]{\begingroup\@sanitize@url \@url }%
\providecommand \@url [1]{\endgroup\@href {#1}{\urlprefix }}%
\providecommand \urlprefix  [0]{URL }%
\providecommand \Eprint [0]{\href }%
\providecommand \doibase [0]{http://dx.doi.org/}%
\providecommand \selectlanguage [0]{\@gobble}%
\providecommand \bibinfo  [0]{\@secondoftwo}%
\providecommand \bibfield  [0]{\@secondoftwo}%
\providecommand \translation [1]{[#1]}%
\providecommand \BibitemOpen [0]{}%
\providecommand \bibitemStop [0]{}%
\providecommand \bibitemNoStop [0]{.\EOS\space}%
\providecommand \EOS [0]{\spacefactor3000\relax}%
\providecommand \BibitemShut  [1]{\csname bibitem#1\endcsname}%
\let\auto@bib@innerbib\@empty
\bibitem [{\citenamefont {Aguilar}\ \emph {et~al.}(2019)\citenamefont {Aguilar} \emph {et~al.}}]{Aguilar:2019teb}%
  \BibitemOpen
  \bibfield  {author} {\bibinfo {author} {\bibfnamefont {Arlene~C.}\ \bibnamefont {Aguilar}} \emph {et~al.},\ }\bibfield  {title} {\enquote {\bibinfo {title} {{Pion and Kaon Structure at the Electron-Ion Collider}},}\ }\href {\doibase 10.1140/epja/i2019-12885-0} {\bibfield  {journal} {\bibinfo  {journal} {Eur. Phys. J. A}\ }\textbf {\bibinfo {volume} {55}},\ \bibinfo {pages} {190} (\bibinfo {year} {2019})},\ \Eprint {http://arxiv.org/abs/1907.08218} {arXiv:1907.08218 [nucl-ex]} \BibitemShut {NoStop}%
\bibitem [{\citenamefont {Abdul~Khalek}\ \emph {et~al.}(2022)\citenamefont {Abdul~Khalek} \emph {et~al.}}]{AbdulKhalek:2021gbh}%
  \BibitemOpen
  \bibfield  {author} {\bibinfo {author} {\bibfnamefont {R.}~\bibnamefont {Abdul~Khalek}} \emph {et~al.},\ }\bibfield  {title} {\enquote {\bibinfo {title} {{Science Requirements and Detector Concepts for the Electron-Ion Collider}: {EIC Yellow Report}},}\ }\href {\doibase 10.1016/j.nuclphysa.2022.122447} {\bibfield  {journal} {\bibinfo  {journal} {Nucl. Phys. A}\ }\textbf {\bibinfo {volume} {1026}},\ \bibinfo {pages} {122447} (\bibinfo {year} {2022})},\ \Eprint {http://arxiv.org/abs/2103.05419} {arXiv:2103.05419 [physics.ins-det]} \BibitemShut {NoStop}%
\bibitem [{\citenamefont {Xie}\ \emph {et~al.}(2022)\citenamefont {Xie}, \citenamefont {Han}, \citenamefont {Wang},\ and\ \citenamefont {Chen}}]{Xie:2021ypc}%
  \BibitemOpen
  \bibfield  {author} {\bibinfo {author} {\bibfnamefont {Gang}\ \bibnamefont {Xie}}, \bibinfo {author} {\bibfnamefont {Chengdong}\ \bibnamefont {Han}}, \bibinfo {author} {\bibfnamefont {Rong}\ \bibnamefont {Wang}}, \ and\ \bibinfo {author} {\bibfnamefont {Xurong}\ \bibnamefont {Chen}},\ }\bibfield  {title} {\enquote {\bibinfo {title} {{Tackling the kaon structure function at EicC *}},}\ }\href {\doibase 10.1088/1674-1137/ac5b0e} {\bibfield  {journal} {\bibinfo  {journal} {Chin. Phys. C}\ }\textbf {\bibinfo {volume} {46}},\ \bibinfo {pages} {064107} (\bibinfo {year} {2022})},\ \Eprint {http://arxiv.org/abs/2109.08483} {arXiv:2109.08483 [hep-ph]} \BibitemShut {NoStop}%
\bibitem [{\citenamefont {Conway}\ \emph {et~al.}(1989)\citenamefont {Conway} \emph {et~al.}}]{Conway:1989fs}%
  \BibitemOpen
  \bibfield  {author} {\bibinfo {author} {\bibfnamefont {J.~S.}\ \bibnamefont {Conway}} \emph {et~al.},\ }\bibfield  {title} {\enquote {\bibinfo {title} {{Experimental Study of Muon Pairs Produced by 252-GeV Pions on Tungsten}},}\ }\href {\doibase 10.1103/PhysRevD.39.92} {\bibfield  {journal} {\bibinfo  {journal} {Phys. Rev. D}\ }\textbf {\bibinfo {volume} {39}},\ \bibinfo {pages} {92--122} (\bibinfo {year} {1989})}\BibitemShut {NoStop}%
\bibitem [{\citenamefont {Badier}\ \emph {et~al.}(1980)\citenamefont {Badier} \emph {et~al.}}]{Saclay-CERN-CollegedeFrance-EcolePoly-Orsay:1980fhh}%
  \BibitemOpen
  \bibfield  {author} {\bibinfo {author} {\bibfnamefont {J.}~\bibnamefont {Badier}} \emph {et~al.} (\bibinfo {collaboration} {Saclay-CERN-College de France-Ecole Poly-Orsay}),\ }\bibfield  {title} {\enquote {\bibinfo {title} {{Measurement of the $K^- / \pi^-$ Structure Function Ratio Using the {Drell-Yan} Process}},}\ }\href {\doibase 10.1016/0370-2693(80)90530-4} {\bibfield  {journal} {\bibinfo  {journal} {Phys. Lett. B}\ }\textbf {\bibinfo {volume} {93}},\ \bibinfo {pages} {354--356} (\bibinfo {year} {1980})}\BibitemShut {NoStop}%
\bibitem [{\citenamefont {Barry}\ \emph {et~al.}(2018)\citenamefont {Barry}, \citenamefont {Sato}, \citenamefont {Melnitchouk},\ and\ \citenamefont {Ji}}]{Barry:2018ort}%
  \BibitemOpen
  \bibfield  {author} {\bibinfo {author} {\bibfnamefont {P.~C.}\ \bibnamefont {Barry}}, \bibinfo {author} {\bibfnamefont {N.}~\bibnamefont {Sato}}, \bibinfo {author} {\bibfnamefont {W.}~\bibnamefont {Melnitchouk}}, \ and\ \bibinfo {author} {\bibfnamefont {Chueng-Ryong}\ \bibnamefont {Ji}},\ }\bibfield  {title} {\enquote {\bibinfo {title} {{First Monte Carlo Global QCD Analysis of Pion Parton Distributions}},}\ }\href {\doibase 10.1103/PhysRevLett.121.152001} {\bibfield  {journal} {\bibinfo  {journal} {Phys. Rev. Lett.}\ }\textbf {\bibinfo {volume} {121}},\ \bibinfo {pages} {152001} (\bibinfo {year} {2018})},\ \Eprint {http://arxiv.org/abs/1804.01965} {arXiv:1804.01965 [hep-ph]} \BibitemShut {NoStop}%
\bibitem [{\citenamefont {Novikov}\ \emph {et~al.}(2020)\citenamefont {Novikov} \emph {et~al.}}]{Novikov:2020snp}%
  \BibitemOpen
  \bibfield  {author} {\bibinfo {author} {\bibfnamefont {Ivan}\ \bibnamefont {Novikov}} \emph {et~al.},\ }\bibfield  {title} {\enquote {\bibinfo {title} {{Parton Distribution Functions of the Charged Pion Within The xFitter Framework}},}\ }\href {\doibase 10.1103/PhysRevD.102.014040} {\bibfield  {journal} {\bibinfo  {journal} {Phys. Rev. D}\ }\textbf {\bibinfo {volume} {102}},\ \bibinfo {pages} {014040} (\bibinfo {year} {2020})},\ \Eprint {http://arxiv.org/abs/2002.02902} {arXiv:2002.02902 [hep-ph]} \BibitemShut {NoStop}%
\bibitem [{\citenamefont {Barry}\ \emph {et~al.}(2021)\citenamefont {Barry}, \citenamefont {Ji}, \citenamefont {Sato},\ and\ \citenamefont {Melnitchouk}}]{Barry:2021osv}%
  \BibitemOpen
  \bibfield  {author} {\bibinfo {author} {\bibfnamefont {P.~C.}\ \bibnamefont {Barry}}, \bibinfo {author} {\bibfnamefont {Chueng-Ryong}\ \bibnamefont {Ji}}, \bibinfo {author} {\bibfnamefont {N.}~\bibnamefont {Sato}}, \ and\ \bibinfo {author} {\bibfnamefont {W.}~\bibnamefont {Melnitchouk}} (\bibinfo {collaboration} {Jefferson Lab Angular Momentum (JAM)}),\ }\bibfield  {title} {\enquote {\bibinfo {title} {{Global QCD Analysis of Pion Parton Distributions with Threshold Resummation}},}\ }\href {\doibase 10.1103/PhysRevLett.127.232001} {\bibfield  {journal} {\bibinfo  {journal} {Phys. Rev. Lett.}\ }\textbf {\bibinfo {volume} {127}},\ \bibinfo {pages} {232001} (\bibinfo {year} {2021})},\ \Eprint {http://arxiv.org/abs/2108.05822} {arXiv:2108.05822 [hep-ph]} \BibitemShut {NoStop}%
\bibitem [{\citenamefont {Barry}\ \emph {et~al.}(2022)\citenamefont {Barry} \emph {et~al.}}]{JeffersonLabAngularMomentumJAM:2022aix}%
  \BibitemOpen
  \bibfield  {author} {\bibinfo {author} {\bibfnamefont {P.~C.}\ \bibnamefont {Barry}} \emph {et~al.} (\bibinfo {collaboration} {Jefferson Lab Angular Momentum (JAM), HadStruc}),\ }\bibfield  {title} {\enquote {\bibinfo {title} {{Complementarity of experimental and lattice QCD data on pion parton distributions}},}\ }\href {\doibase 10.1103/PhysRevD.105.114051} {\bibfield  {journal} {\bibinfo  {journal} {Phys. Rev. D}\ }\textbf {\bibinfo {volume} {105}},\ \bibinfo {pages} {114051} (\bibinfo {year} {2022})},\ \Eprint {http://arxiv.org/abs/2204.00543} {arXiv:2204.00543 [hep-ph]} \BibitemShut {NoStop}%
\bibitem [{\citenamefont {Kotz}\ \emph {et~al.}(2023)\citenamefont {Kotz}, \citenamefont {Courtoy}, \citenamefont {Nadolsky}, \citenamefont {Olness},\ and\ \citenamefont {Ponce-Chavez}}]{Kotz:2023pbu}%
  \BibitemOpen
  \bibfield  {author} {\bibinfo {author} {\bibfnamefont {Lucas}\ \bibnamefont {Kotz}}, \bibinfo {author} {\bibfnamefont {Aurore}\ \bibnamefont {Courtoy}}, \bibinfo {author} {\bibfnamefont {Pavel}\ \bibnamefont {Nadolsky}}, \bibinfo {author} {\bibfnamefont {Fredrick}\ \bibnamefont {Olness}}, \ and\ \bibinfo {author} {\bibfnamefont {Maximiliano}\ \bibnamefont {Ponce-Chavez}},\ }\bibfield  {title} {\enquote {\bibinfo {title} {{An analysis of parton distributions in a pion with B\'ezier parametrizations}},}\ }\href@noop {} {\  (\bibinfo {year} {2023})},\ \Eprint {http://arxiv.org/abs/2311.08447} {arXiv:2311.08447 [hep-ph]} \BibitemShut {NoStop}%
\bibitem [{\citenamefont {Chen}\ \emph {et~al.}(2016)\citenamefont {Chen}, \citenamefont {Chang}, \citenamefont {Roberts}, \citenamefont {Wan},\ and\ \citenamefont {Zong}}]{Chen:2016sno}%
  \BibitemOpen
  \bibfield  {author} {\bibinfo {author} {\bibfnamefont {Chen}\ \bibnamefont {Chen}}, \bibinfo {author} {\bibfnamefont {Lei}\ \bibnamefont {Chang}}, \bibinfo {author} {\bibfnamefont {Craig~D.}\ \bibnamefont {Roberts}}, \bibinfo {author} {\bibfnamefont {Shaolong}\ \bibnamefont {Wan}}, \ and\ \bibinfo {author} {\bibfnamefont {Hong-Shi}\ \bibnamefont {Zong}},\ }\bibfield  {title} {\enquote {\bibinfo {title} {{Valence-quark distribution functions in the kaon and pion}},}\ }\href {\doibase 10.1103/PhysRevD.93.074021} {\bibfield  {journal} {\bibinfo  {journal} {Phys. Rev. D}\ }\textbf {\bibinfo {volume} {93}},\ \bibinfo {pages} {074021} (\bibinfo {year} {2016})},\ \Eprint {http://arxiv.org/abs/1602.01502} {arXiv:1602.01502 [nucl-th]} \BibitemShut {NoStop}%
\bibitem [{\citenamefont {Shi}\ \emph {et~al.}(2018)\citenamefont {Shi}, \citenamefont {Mezrag},\ and\ \citenamefont {Zong}}]{Shi:2018mcb}%
  \BibitemOpen
  \bibfield  {author} {\bibinfo {author} {\bibfnamefont {Chao}\ \bibnamefont {Shi}}, \bibinfo {author} {\bibfnamefont {C\'edric}\ \bibnamefont {Mezrag}}, \ and\ \bibinfo {author} {\bibfnamefont {Hong-shi}\ \bibnamefont {Zong}},\ }\bibfield  {title} {\enquote {\bibinfo {title} {{Pion and kaon valence quark distribution functions from Dyson-Schwinger equations}},}\ }\href {\doibase 10.1103/PhysRevD.98.054029} {\bibfield  {journal} {\bibinfo  {journal} {Phys. Rev. D}\ }\textbf {\bibinfo {volume} {98}},\ \bibinfo {pages} {054029} (\bibinfo {year} {2018})},\ \Eprint {http://arxiv.org/abs/1806.10232} {arXiv:1806.10232 [nucl-th]} \BibitemShut {NoStop}%
\bibitem [{\citenamefont {Lan}\ \emph {et~al.}(2020)\citenamefont {Lan}, \citenamefont {Mondal}, \citenamefont {Jia}, \citenamefont {Zhao},\ and\ \citenamefont {Vary}}]{Lan:2019rba}%
  \BibitemOpen
  \bibfield  {author} {\bibinfo {author} {\bibfnamefont {Jiangshan}\ \bibnamefont {Lan}}, \bibinfo {author} {\bibfnamefont {Chandan}\ \bibnamefont {Mondal}}, \bibinfo {author} {\bibfnamefont {Shaoyang}\ \bibnamefont {Jia}}, \bibinfo {author} {\bibfnamefont {Xingbo}\ \bibnamefont {Zhao}}, \ and\ \bibinfo {author} {\bibfnamefont {James~P.}\ \bibnamefont {Vary}},\ }\bibfield  {title} {\enquote {\bibinfo {title} {{Pion and kaon parton distribution functions from basis light front quantization and QCD evolution}},}\ }\href {\doibase 10.1103/PhysRevD.101.034024} {\bibfield  {journal} {\bibinfo  {journal} {Phys. Rev. D}\ }\textbf {\bibinfo {volume} {101}},\ \bibinfo {pages} {034024} (\bibinfo {year} {2020})},\ \Eprint {http://arxiv.org/abs/1907.01509} {arXiv:1907.01509 [nucl-th]} \BibitemShut {NoStop}%
\bibitem [{\citenamefont {Bednar}\ \emph {et~al.}(2020)\citenamefont {Bednar}, \citenamefont {Clo\"et},\ and\ \citenamefont {Tandy}}]{Bednar:2018mtf}%
  \BibitemOpen
  \bibfield  {author} {\bibinfo {author} {\bibfnamefont {Kyle~D.}\ \bibnamefont {Bednar}}, \bibinfo {author} {\bibfnamefont {Ian~C.}\ \bibnamefont {Clo\"et}}, \ and\ \bibinfo {author} {\bibfnamefont {Peter~C.}\ \bibnamefont {Tandy}},\ }\bibfield  {title} {\enquote {\bibinfo {title} {{Distinguishing Quarks and Gluons in Pion and Kaon Parton Distribution Functions}},}\ }\href {\doibase 10.1103/PhysRevLett.124.042002} {\bibfield  {journal} {\bibinfo  {journal} {Phys. Rev. Lett.}\ }\textbf {\bibinfo {volume} {124}},\ \bibinfo {pages} {042002} (\bibinfo {year} {2020})},\ \Eprint {http://arxiv.org/abs/1811.12310} {arXiv:1811.12310 [nucl-th]} \BibitemShut {NoStop}%
\bibitem [{\citenamefont {Cui}\ \emph {et~al.}(2020)\citenamefont {Cui}, \citenamefont {Ding}, \citenamefont {Gao}, \citenamefont {Raya}, \citenamefont {Binosi}, \citenamefont {Chang}, \citenamefont {Roberts}, \citenamefont {Rodr\'\i{}guez-Quintero},\ and\ \citenamefont {Schmidt}}]{Cui:2020tdf}%
  \BibitemOpen
  \bibfield  {author} {\bibinfo {author} {\bibfnamefont {Zhu-Fang}\ \bibnamefont {Cui}}, \bibinfo {author} {\bibfnamefont {Minghui}\ \bibnamefont {Ding}}, \bibinfo {author} {\bibfnamefont {Fei}\ \bibnamefont {Gao}}, \bibinfo {author} {\bibfnamefont {Kh\'epani}\ \bibnamefont {Raya}}, \bibinfo {author} {\bibfnamefont {Daniele}\ \bibnamefont {Binosi}}, \bibinfo {author} {\bibfnamefont {Lei}\ \bibnamefont {Chang}}, \bibinfo {author} {\bibfnamefont {Craig~D}\ \bibnamefont {Roberts}}, \bibinfo {author} {\bibfnamefont {Jose}\ \bibnamefont {Rodr\'\i{}guez-Quintero}}, \ and\ \bibinfo {author} {\bibfnamefont {Sebastian~M}\ \bibnamefont {Schmidt}},\ }\bibfield  {title} {\enquote {\bibinfo {title} {{Kaon and pion parton distributions}},}\ }\href {\doibase 10.1140/epjc/s10052-020-08578-4} {\bibfield  {journal} {\bibinfo  {journal} {Eur. Phys. J. C}\ }\textbf {\bibinfo {volume} {80}},\ \bibinfo {pages} {1064} (\bibinfo {year} {2020})}\BibitemShut {NoStop}%
\bibitem [{\citenamefont {Han}\ \emph {et~al.}(2021)\citenamefont {Han}, \citenamefont {Xie}, \citenamefont {Wang},\ and\ \citenamefont {Chen}}]{Han:2020vjp}%
  \BibitemOpen
  \bibfield  {author} {\bibinfo {author} {\bibfnamefont {Chengdong}\ \bibnamefont {Han}}, \bibinfo {author} {\bibfnamefont {Gang}\ \bibnamefont {Xie}}, \bibinfo {author} {\bibfnamefont {Rong}\ \bibnamefont {Wang}}, \ and\ \bibinfo {author} {\bibfnamefont {Xurong}\ \bibnamefont {Chen}},\ }\bibfield  {title} {\enquote {\bibinfo {title} {{An Analysis of Parton Distribution Functions of the Pion and the Kaon with the Maximum Entropy Input}},}\ }\href {\doibase 10.1140/epjc/s10052-021-09087-8} {\bibfield  {journal} {\bibinfo  {journal} {Eur. Phys. J. C}\ }\textbf {\bibinfo {volume} {81}},\ \bibinfo {pages} {302} (\bibinfo {year} {2021})},\ \Eprint {http://arxiv.org/abs/2010.14284} {arXiv:2010.14284 [hep-ph]} \BibitemShut {NoStop}%
\bibitem [{\citenamefont {Roberts}\ \emph {et~al.}(2021)\citenamefont {Roberts}, \citenamefont {Richards}, \citenamefont {Horn},\ and\ \citenamefont {Chang}}]{Roberts:2021nhw}%
  \BibitemOpen
  \bibfield  {author} {\bibinfo {author} {\bibfnamefont {Craig~D.}\ \bibnamefont {Roberts}}, \bibinfo {author} {\bibfnamefont {David~G.}\ \bibnamefont {Richards}}, \bibinfo {author} {\bibfnamefont {Tanja}\ \bibnamefont {Horn}}, \ and\ \bibinfo {author} {\bibfnamefont {Lei}\ \bibnamefont {Chang}},\ }\bibfield  {title} {\enquote {\bibinfo {title} {{Insights into the emergence of mass from studies of pion and kaon structure}},}\ }\href {\doibase 10.1016/j.ppnp.2021.103883} {\bibfield  {journal} {\bibinfo  {journal} {Prog. Part. Nucl. Phys.}\ }\textbf {\bibinfo {volume} {120}},\ \bibinfo {pages} {103883} (\bibinfo {year} {2021})},\ \Eprint {http://arxiv.org/abs/2102.01765} {arXiv:2102.01765 [hep-ph]} \BibitemShut {NoStop}%
\bibitem [{\citenamefont {Pasquini}\ \emph {et~al.}(2023)\citenamefont {Pasquini}, \citenamefont {Rodini},\ and\ \citenamefont {Venturini}}]{Pasquini:2023aaf}%
  \BibitemOpen
  \bibfield  {author} {\bibinfo {author} {\bibfnamefont {Barbara}\ \bibnamefont {Pasquini}}, \bibinfo {author} {\bibfnamefont {Simone}\ \bibnamefont {Rodini}}, \ and\ \bibinfo {author} {\bibfnamefont {Simone}\ \bibnamefont {Venturini}} (\bibinfo {collaboration} {MAP (Multi-dimensional Analyses of Partonic distributions)}),\ }\bibfield  {title} {\enquote {\bibinfo {title} {{Valence quark, sea, and gluon content of the pion from the parton distribution functions and the electromagnetic form factor}},}\ }\href {\doibase 10.1103/PhysRevD.107.114023} {\bibfield  {journal} {\bibinfo  {journal} {Phys. Rev. D}\ }\textbf {\bibinfo {volume} {107}},\ \bibinfo {pages} {114023} (\bibinfo {year} {2023})},\ \Eprint {http://arxiv.org/abs/2303.01789} {arXiv:2303.01789 [hep-ph]} \BibitemShut {NoStop}%
\bibitem [{\citenamefont {Constantinou}(2021)}]{Constantinou:2020pek}%
  \BibitemOpen
  \bibfield  {author} {\bibinfo {author} {\bibfnamefont {Martha}\ \bibnamefont {Constantinou}},\ }\bibfield  {title} {\enquote {\bibinfo {title} {{The x-dependence of hadronic parton distributions: A review on the progress of lattice QCD}},}\ }\href {\doibase 10.1140/epja/s10050-021-00353-7} {\bibfield  {journal} {\bibinfo  {journal} {Eur. Phys. J. A}\ }\textbf {\bibinfo {volume} {57}},\ \bibinfo {pages} {77} (\bibinfo {year} {2021})},\ \Eprint {http://arxiv.org/abs/2010.02445} {arXiv:2010.02445 [hep-lat]} \BibitemShut {NoStop}%
\bibitem [{\citenamefont {Martinelli}\ and\ \citenamefont {Sachrajda}(1987)}]{Martinelli:1987zd}%
  \BibitemOpen
  \bibfield  {author} {\bibinfo {author} {\bibfnamefont {G.}~\bibnamefont {Martinelli}}\ and\ \bibinfo {author} {\bibfnamefont {Christopher~T.}\ \bibnamefont {Sachrajda}},\ }\bibfield  {title} {\enquote {\bibinfo {title} {{Pion Structure Functions From Lattice {QCD}}},}\ }\href {\doibase 10.1016/0370-2693(87)90601-0} {\bibfield  {journal} {\bibinfo  {journal} {Phys. Lett. B}\ }\textbf {\bibinfo {volume} {196}},\ \bibinfo {pages} {184--190} (\bibinfo {year} {1987})}\BibitemShut {NoStop}%
\bibitem [{\citenamefont {Best}\ \emph {et~al.}(1997)\citenamefont {Best}, \citenamefont {Göckeler}, \citenamefont {Horsley}, \citenamefont {Ilgenfritz}, \citenamefont {Perlt}, \citenamefont {Rakow}, \citenamefont {Schäfer}, \citenamefont {Schierholz}, \citenamefont {Schiller},\ and\ \citenamefont {Schramm}}]{Best:1997qp}%
  \BibitemOpen
  \bibfield  {author} {\bibinfo {author} {\bibfnamefont {C.}~\bibnamefont {Best}}, \bibinfo {author} {\bibfnamefont {M.}~\bibnamefont {Göckeler}}, \bibinfo {author} {\bibfnamefont {R.}~\bibnamefont {Horsley}}, \bibinfo {author} {\bibfnamefont {Ernst-Michael}\ \bibnamefont {Ilgenfritz}}, \bibinfo {author} {\bibfnamefont {H.}~\bibnamefont {Perlt}}, \bibinfo {author} {\bibfnamefont {Paul E.~L.}\ \bibnamefont {Rakow}}, \bibinfo {author} {\bibfnamefont {A.}~\bibnamefont {Schäfer}}, \bibinfo {author} {\bibfnamefont {G.}~\bibnamefont {Schierholz}}, \bibinfo {author} {\bibfnamefont {A.}~\bibnamefont {Schiller}}, \ and\ \bibinfo {author} {\bibfnamefont {S.}~\bibnamefont {Schramm}},\ }\bibfield  {title} {\enquote {\bibinfo {title} {{Pion and rho structure functions from lattice QCD}},}\ }\href {\doibase 10.1103/PhysRevD.56.2743} {\bibfield  {journal} {\bibinfo  {journal} {Phys. Rev. D}\ }\textbf {\bibinfo {volume} {56}},\ \bibinfo {pages} {2743--2754} (\bibinfo {year} {1997})},\ \Eprint
  {http://arxiv.org/abs/hep-lat/9703014} {arXiv:hep-lat/9703014} \BibitemShut {NoStop}%
\bibitem [{\citenamefont {Guagnelli}\ \emph {et~al.}(2005)\citenamefont {Guagnelli}, \citenamefont {Jansen}, \citenamefont {Palombi}, \citenamefont {Petronzio}, \citenamefont {Shindler},\ and\ \citenamefont {Wetzorke}}]{Guagnelli:2004ga}%
  \BibitemOpen
  \bibfield  {author} {\bibinfo {author} {\bibfnamefont {M.}~\bibnamefont {Guagnelli}}, \bibinfo {author} {\bibfnamefont {K.}~\bibnamefont {Jansen}}, \bibinfo {author} {\bibfnamefont {F.}~\bibnamefont {Palombi}}, \bibinfo {author} {\bibfnamefont {R.}~\bibnamefont {Petronzio}}, \bibinfo {author} {\bibfnamefont {A.}~\bibnamefont {Shindler}}, \ and\ \bibinfo {author} {\bibfnamefont {I.}~\bibnamefont {Wetzorke}} (\bibinfo {collaboration} {Zeuthen-Rome (ZeRo)}),\ }\bibfield  {title} {\enquote {\bibinfo {title} {{Non-perturbative pion matrix element of a twist-2 operator from the lattice}},}\ }\href {\doibase 10.1140/epjc/s2005-02121-5} {\bibfield  {journal} {\bibinfo  {journal} {Eur. Phys. J. C}\ }\textbf {\bibinfo {volume} {40}},\ \bibinfo {pages} {69--80} (\bibinfo {year} {2005})},\ \Eprint {http://arxiv.org/abs/hep-lat/0405027} {arXiv:hep-lat/0405027} \BibitemShut {NoStop}%
\bibitem [{\citenamefont {Capitani}\ \emph {et~al.}(2006)\citenamefont {Capitani}, \citenamefont {Jansen}, \citenamefont {Papinutto}, \citenamefont {Shindler}, \citenamefont {Urbach},\ and\ \citenamefont {Wetzorke}}]{Capitani:2005jp}%
  \BibitemOpen
  \bibfield  {author} {\bibinfo {author} {\bibfnamefont {S.}~\bibnamefont {Capitani}}, \bibinfo {author} {\bibfnamefont {K.}~\bibnamefont {Jansen}}, \bibinfo {author} {\bibfnamefont {M.}~\bibnamefont {Papinutto}}, \bibinfo {author} {\bibfnamefont {A.}~\bibnamefont {Shindler}}, \bibinfo {author} {\bibfnamefont {C.}~\bibnamefont {Urbach}}, \ and\ \bibinfo {author} {\bibfnamefont {I.}~\bibnamefont {Wetzorke}},\ }\bibfield  {title} {\enquote {\bibinfo {title} {{Parton distribution functions with twisted mass fermions}},}\ }\href {\doibase 10.1016/j.physletb.2006.02.047} {\bibfield  {journal} {\bibinfo  {journal} {Phys. Lett. B}\ }\textbf {\bibinfo {volume} {639}},\ \bibinfo {pages} {520--526} (\bibinfo {year} {2006})},\ \Eprint {http://arxiv.org/abs/hep-lat/0511013} {arXiv:hep-lat/0511013} \BibitemShut {NoStop}%
\bibitem [{\citenamefont {Abdel-Rehim}\ \emph {et~al.}(2015)\citenamefont {Abdel-Rehim} \emph {et~al.}}]{Abdel-Rehim:2015owa}%
  \BibitemOpen
  \bibfield  {author} {\bibinfo {author} {\bibfnamefont {A.}~\bibnamefont {Abdel-Rehim}} \emph {et~al.},\ }\bibfield  {title} {\enquote {\bibinfo {title} {{Nucleon and pion structure with lattice QCD simulations at physical value of the pion mass}},}\ }\href {\doibase 10.1103/PhysRevD.92.114513} {\bibfield  {journal} {\bibinfo  {journal} {Phys. Rev. D}\ }\textbf {\bibinfo {volume} {92}},\ \bibinfo {pages} {114513} (\bibinfo {year} {2015})},\ \bibinfo {note} {[Erratum: Phys.Rev.D 93, 039904 (2016)]},\ \Eprint {http://arxiv.org/abs/1507.04936} {arXiv:1507.04936 [hep-lat]} \BibitemShut {NoStop}%
\bibitem [{\citenamefont {Oehm}\ \emph {et~al.}(2019)\citenamefont {Oehm}, \citenamefont {Alexandrou}, \citenamefont {Constantinou}, \citenamefont {Jansen}, \citenamefont {Koutsou}, \citenamefont {Kostrzewa}, \citenamefont {Steffens}, \citenamefont {Urbach},\ and\ \citenamefont {Zafeiropoulos}}]{Oehm:2018jvm}%
  \BibitemOpen
  \bibfield  {author} {\bibinfo {author} {\bibfnamefont {M.}~\bibnamefont {Oehm}}, \bibinfo {author} {\bibfnamefont {C.}~\bibnamefont {Alexandrou}}, \bibinfo {author} {\bibfnamefont {M.}~\bibnamefont {Constantinou}}, \bibinfo {author} {\bibfnamefont {K.}~\bibnamefont {Jansen}}, \bibinfo {author} {\bibfnamefont {G.}~\bibnamefont {Koutsou}}, \bibinfo {author} {\bibfnamefont {B.}~\bibnamefont {Kostrzewa}}, \bibinfo {author} {\bibfnamefont {F.}~\bibnamefont {Steffens}}, \bibinfo {author} {\bibfnamefont {C.}~\bibnamefont {Urbach}}, \ and\ \bibinfo {author} {\bibfnamefont {S.}~\bibnamefont {Zafeiropoulos}},\ }\bibfield  {title} {\enquote {\bibinfo {title} {{$\langle x\rangle$ and $\langle x^2\rangle$ of the pion PDF from lattice QCD with $N_f=2+1+1$ dynamical quark flavors}},}\ }\href {\doibase 10.1103/PhysRevD.99.014508} {\bibfield  {journal} {\bibinfo  {journal} {Phys. Rev. D}\ }\textbf {\bibinfo {volume} {99}},\ \bibinfo {pages} {014508} (\bibinfo {year} {2019})},\ \Eprint {http://arxiv.org/abs/1810.09743}
  {arXiv:1810.09743 [hep-lat]} \BibitemShut {NoStop}%
\bibitem [{\citenamefont {Alexandrou}\ \emph {et~al.}(2021{\natexlab{a}})\citenamefont {Alexandrou}, \citenamefont {Bacchio}, \citenamefont {Cloet}, \citenamefont {Constantinou}, \citenamefont {Hadjiyiannakou}, \citenamefont {Koutsou},\ and\ \citenamefont {Lauer}}]{Alexandrou:2020gxs}%
  \BibitemOpen
  \bibfield  {author} {\bibinfo {author} {\bibfnamefont {Constantia}\ \bibnamefont {Alexandrou}}, \bibinfo {author} {\bibfnamefont {Simone}\ \bibnamefont {Bacchio}}, \bibinfo {author} {\bibfnamefont {Ian}\ \bibnamefont {Cloet}}, \bibinfo {author} {\bibfnamefont {Martha}\ \bibnamefont {Constantinou}}, \bibinfo {author} {\bibfnamefont {Kyriakos}\ \bibnamefont {Hadjiyiannakou}}, \bibinfo {author} {\bibfnamefont {Giannis}\ \bibnamefont {Koutsou}}, \ and\ \bibinfo {author} {\bibfnamefont {Colin}\ \bibnamefont {Lauer}} (\bibinfo {collaboration} {ETM}),\ }\bibfield  {title} {\enquote {\bibinfo {title} {{Mellin moments $\langle x \rangle$ and $\langle x^2 \rangle$ for the pion and kaon from lattice QCD}},}\ }\href {\doibase 10.1103/PhysRevD.103.014508} {\bibfield  {journal} {\bibinfo  {journal} {Phys. Rev. D}\ }\textbf {\bibinfo {volume} {103}},\ \bibinfo {pages} {014508} (\bibinfo {year} {2021}{\natexlab{a}})},\ \Eprint {http://arxiv.org/abs/2010.03495} {arXiv:2010.03495 [hep-lat]} \BibitemShut {NoStop}%
\bibitem [{\citenamefont {L\"offler}\ \emph {et~al.}(2022)\citenamefont {L\"offler}, \citenamefont {Wein}, \citenamefont {Wurm}, \citenamefont {Weish\"aupl}, \citenamefont {Jenkins}, \citenamefont {R\"odl}, \citenamefont {Sch\"afer},\ and\ \citenamefont {Walter}}]{Loffler:2021afv}%
  \BibitemOpen
  \bibfield  {author} {\bibinfo {author} {\bibfnamefont {Marius}\ \bibnamefont {L\"offler}}, \bibinfo {author} {\bibfnamefont {Philipp}\ \bibnamefont {Wein}}, \bibinfo {author} {\bibfnamefont {Thomas}\ \bibnamefont {Wurm}}, \bibinfo {author} {\bibfnamefont {Simon}\ \bibnamefont {Weish\"aupl}}, \bibinfo {author} {\bibfnamefont {Daniel}\ \bibnamefont {Jenkins}}, \bibinfo {author} {\bibfnamefont {Rudolf}\ \bibnamefont {R\"odl}}, \bibinfo {author} {\bibfnamefont {Andreas}\ \bibnamefont {Sch\"afer}}, \ and\ \bibinfo {author} {\bibfnamefont {Lisa}\ \bibnamefont {Walter}} (\bibinfo {collaboration} {RQCD}),\ }\bibfield  {title} {\enquote {\bibinfo {title} {{Mellin moments of spin dependent and independent PDFs of the pion and rho meson}},}\ }\href {\doibase 10.1103/PhysRevD.105.014505} {\bibfield  {journal} {\bibinfo  {journal} {Phys. Rev. D}\ }\textbf {\bibinfo {volume} {105}},\ \bibinfo {pages} {014505} (\bibinfo {year} {2022})},\ \Eprint {http://arxiv.org/abs/2108.07544} {arXiv:2108.07544 [hep-lat]} \BibitemShut
  {NoStop}%
\bibitem [{\citenamefont {Alexandrou}\ \emph {et~al.}(2021{\natexlab{b}})\citenamefont {Alexandrou} \emph {et~al.}}]{ExtendedTwistedMass:2021rdx}%
  \BibitemOpen
  \bibfield  {author} {\bibinfo {author} {\bibfnamefont {Constantia}\ \bibnamefont {Alexandrou}} \emph {et~al.} (\bibinfo {collaboration} {Extended Twisted Mass}),\ }\bibfield  {title} {\enquote {\bibinfo {title} {{Quark and Gluon Momentum Fractions in the Pion from Nf=2+1+1 Lattice QCD}},}\ }\href {\doibase 10.1103/PhysRevLett.127.252001} {\bibfield  {journal} {\bibinfo  {journal} {Phys. Rev. Lett.}\ }\textbf {\bibinfo {volume} {127}},\ \bibinfo {pages} {252001} (\bibinfo {year} {2021}{\natexlab{b}})},\ \Eprint {http://arxiv.org/abs/2109.10692} {arXiv:2109.10692 [hep-lat]} \BibitemShut {NoStop}%
\bibitem [{\citenamefont {Hackett}\ \emph {et~al.}(2023)\citenamefont {Hackett}, \citenamefont {Oare}, \citenamefont {Pefkou},\ and\ \citenamefont {Shanahan}}]{Hackett:2023nkr}%
  \BibitemOpen
  \bibfield  {author} {\bibinfo {author} {\bibfnamefont {Daniel~C.}\ \bibnamefont {Hackett}}, \bibinfo {author} {\bibfnamefont {Patrick~R.}\ \bibnamefont {Oare}}, \bibinfo {author} {\bibfnamefont {Dimitra~A.}\ \bibnamefont {Pefkou}}, \ and\ \bibinfo {author} {\bibfnamefont {Phiala~E.}\ \bibnamefont {Shanahan}},\ }\bibfield  {title} {\enquote {\bibinfo {title} {{Gravitational form factors of the pion from lattice QCD}},}\ }\href@noop {} {\  (\bibinfo {year} {2023})},\ \Eprint {http://arxiv.org/abs/2307.11707} {arXiv:2307.11707 [hep-lat]} \BibitemShut {NoStop}%
\bibitem [{\citenamefont {Good}\ \emph {et~al.}(2023)\citenamefont {Good}, \citenamefont {Hasan}, \citenamefont {Chevis},\ and\ \citenamefont {Lin}}]{Good:2023ecp}%
  \BibitemOpen
  \bibfield  {author} {\bibinfo {author} {\bibfnamefont {William}\ \bibnamefont {Good}}, \bibinfo {author} {\bibfnamefont {Kinza}\ \bibnamefont {Hasan}}, \bibinfo {author} {\bibfnamefont {Allison}\ \bibnamefont {Chevis}}, \ and\ \bibinfo {author} {\bibfnamefont {Huey-Wen}\ \bibnamefont {Lin}},\ }\bibfield  {title} {\enquote {\bibinfo {title} {{The Gluon Moment and Parton Distribution Function of the Pion from $N_f = 2 + 1 + 1$ Lattice QCD}},}\ }\href@noop {} {\  (\bibinfo {year} {2023})},\ \Eprint {http://arxiv.org/abs/2310.12034} {arXiv:2310.12034 [hep-lat]} \BibitemShut {NoStop}%
\bibitem [{\citenamefont {Zhang}\ \emph {et~al.}(2019)\citenamefont {Zhang}, \citenamefont {Chen}, \citenamefont {Jin}, \citenamefont {Lin}, \citenamefont {Sch\"afer},\ and\ \citenamefont {Zhao}}]{Zhang:2018nsy}%
  \BibitemOpen
  \bibfield  {author} {\bibinfo {author} {\bibfnamefont {Jian-Hui}\ \bibnamefont {Zhang}}, \bibinfo {author} {\bibfnamefont {Jiunn-Wei}\ \bibnamefont {Chen}}, \bibinfo {author} {\bibfnamefont {Luchang}\ \bibnamefont {Jin}}, \bibinfo {author} {\bibfnamefont {Huey-Wen}\ \bibnamefont {Lin}}, \bibinfo {author} {\bibfnamefont {Andreas}\ \bibnamefont {Sch\"afer}}, \ and\ \bibinfo {author} {\bibfnamefont {Yong}\ \bibnamefont {Zhao}},\ }\bibfield  {title} {\enquote {\bibinfo {title} {{First direct lattice-QCD calculation of the $x$-dependence of the pion parton distribution function}},}\ }\href {\doibase 10.1103/PhysRevD.100.034505} {\bibfield  {journal} {\bibinfo  {journal} {Phys. Rev. D}\ }\textbf {\bibinfo {volume} {100}},\ \bibinfo {pages} {034505} (\bibinfo {year} {2019})},\ \Eprint {http://arxiv.org/abs/1804.01483} {arXiv:1804.01483 [hep-lat]} \BibitemShut {NoStop}%
\bibitem [{\citenamefont {Izubuchi}\ \emph {et~al.}(2019)\citenamefont {Izubuchi}, \citenamefont {Jin}, \citenamefont {Kallidonis}, \citenamefont {Karthik}, \citenamefont {Mukherjee}, \citenamefont {Petreczky}, \citenamefont {Shugert},\ and\ \citenamefont {Syritsyn}}]{Izubuchi:2019lyk}%
  \BibitemOpen
  \bibfield  {author} {\bibinfo {author} {\bibfnamefont {Taku}\ \bibnamefont {Izubuchi}}, \bibinfo {author} {\bibfnamefont {Luchang}\ \bibnamefont {Jin}}, \bibinfo {author} {\bibfnamefont {Christos}\ \bibnamefont {Kallidonis}}, \bibinfo {author} {\bibfnamefont {Nikhil}\ \bibnamefont {Karthik}}, \bibinfo {author} {\bibfnamefont {Swagato}\ \bibnamefont {Mukherjee}}, \bibinfo {author} {\bibfnamefont {Peter}\ \bibnamefont {Petreczky}}, \bibinfo {author} {\bibfnamefont {Charles}\ \bibnamefont {Shugert}}, \ and\ \bibinfo {author} {\bibfnamefont {Sergey}\ \bibnamefont {Syritsyn}},\ }\bibfield  {title} {\enquote {\bibinfo {title} {{Valence parton distribution function of pion from fine lattice}},}\ }\href {\doibase 10.1103/PhysRevD.100.034516} {\bibfield  {journal} {\bibinfo  {journal} {Phys. Rev. D}\ }\textbf {\bibinfo {volume} {100}},\ \bibinfo {pages} {034516} (\bibinfo {year} {2019})},\ \Eprint {http://arxiv.org/abs/1905.06349} {arXiv:1905.06349 [hep-lat]} \BibitemShut {NoStop}%
\bibitem [{\citenamefont {Lin}\ \emph {et~al.}(2021)\citenamefont {Lin}, \citenamefont {Chen}, \citenamefont {Fan}, \citenamefont {Zhang},\ and\ \citenamefont {Zhang}}]{Lin:2020ssv}%
  \BibitemOpen
  \bibfield  {author} {\bibinfo {author} {\bibfnamefont {Huey-Wen}\ \bibnamefont {Lin}}, \bibinfo {author} {\bibfnamefont {Jiunn-Wei}\ \bibnamefont {Chen}}, \bibinfo {author} {\bibfnamefont {Zhouyou}\ \bibnamefont {Fan}}, \bibinfo {author} {\bibfnamefont {Jian-Hui}\ \bibnamefont {Zhang}}, \ and\ \bibinfo {author} {\bibfnamefont {Rui}\ \bibnamefont {Zhang}},\ }\bibfield  {title} {\enquote {\bibinfo {title} {{Valence-Quark Distribution of the Kaon and Pion from Lattice QCD}},}\ }\href {\doibase 10.1103/PhysRevD.103.014516} {\bibfield  {journal} {\bibinfo  {journal} {Phys. Rev. D}\ }\textbf {\bibinfo {volume} {103}},\ \bibinfo {pages} {014516} (\bibinfo {year} {2021})},\ \Eprint {http://arxiv.org/abs/2003.14128} {arXiv:2003.14128 [hep-lat]} \BibitemShut {NoStop}%
\bibitem [{\citenamefont {Gao}\ \emph {et~al.}(2020)\citenamefont {Gao}, \citenamefont {Jin}, \citenamefont {Kallidonis}, \citenamefont {Karthik}, \citenamefont {Mukherjee}, \citenamefont {Petreczky}, \citenamefont {Shugert}, \citenamefont {Syritsyn},\ and\ \citenamefont {Zhao}}]{Gao:2020ito}%
  \BibitemOpen
  \bibfield  {author} {\bibinfo {author} {\bibfnamefont {Xiang}\ \bibnamefont {Gao}}, \bibinfo {author} {\bibfnamefont {Luchang}\ \bibnamefont {Jin}}, \bibinfo {author} {\bibfnamefont {Christos}\ \bibnamefont {Kallidonis}}, \bibinfo {author} {\bibfnamefont {Nikhil}\ \bibnamefont {Karthik}}, \bibinfo {author} {\bibfnamefont {Swagato}\ \bibnamefont {Mukherjee}}, \bibinfo {author} {\bibfnamefont {Peter}\ \bibnamefont {Petreczky}}, \bibinfo {author} {\bibfnamefont {Charles}\ \bibnamefont {Shugert}}, \bibinfo {author} {\bibfnamefont {Sergey}\ \bibnamefont {Syritsyn}}, \ and\ \bibinfo {author} {\bibfnamefont {Yong}\ \bibnamefont {Zhao}},\ }\bibfield  {title} {\enquote {\bibinfo {title} {{Valence parton distribution of the pion from lattice QCD: Approaching the continuum limit}},}\ }\href {\doibase 10.1103/PhysRevD.102.094513} {\bibfield  {journal} {\bibinfo  {journal} {Phys. Rev. D}\ }\textbf {\bibinfo {volume} {102}},\ \bibinfo {pages} {094513} (\bibinfo {year} {2020})},\ \Eprint {http://arxiv.org/abs/2007.06590}
  {arXiv:2007.06590 [hep-lat]} \BibitemShut {NoStop}%
\bibitem [{\citenamefont {Jo\'o}\ \emph {et~al.}(2019)\citenamefont {Jo\'o}, \citenamefont {Karpie}, \citenamefont {Orginos}, \citenamefont {Radyushkin}, \citenamefont {Richards}, \citenamefont {Sufian},\ and\ \citenamefont {Zafeiropoulos}}]{Joo:2019bzr}%
  \BibitemOpen
  \bibfield  {author} {\bibinfo {author} {\bibfnamefont {B\'alint}\ \bibnamefont {Jo\'o}}, \bibinfo {author} {\bibfnamefont {Joseph}\ \bibnamefont {Karpie}}, \bibinfo {author} {\bibfnamefont {Kostas}\ \bibnamefont {Orginos}}, \bibinfo {author} {\bibfnamefont {Anatoly~V.}\ \bibnamefont {Radyushkin}}, \bibinfo {author} {\bibfnamefont {David~G.}\ \bibnamefont {Richards}}, \bibinfo {author} {\bibfnamefont {Raza~Sabbir}\ \bibnamefont {Sufian}}, \ and\ \bibinfo {author} {\bibfnamefont {Savvas}\ \bibnamefont {Zafeiropoulos}},\ }\bibfield  {title} {\enquote {\bibinfo {title} {{Pion valence structure from Ioffe-time parton pseudodistribution functions}},}\ }\href {\doibase 10.1103/PhysRevD.100.114512} {\bibfield  {journal} {\bibinfo  {journal} {Phys. Rev. D}\ }\textbf {\bibinfo {volume} {100}},\ \bibinfo {pages} {114512} (\bibinfo {year} {2019})},\ \Eprint {http://arxiv.org/abs/1909.08517} {arXiv:1909.08517 [hep-lat]} \BibitemShut {NoStop}%
\bibitem [{\citenamefont {Salas-Chavira}\ \emph {et~al.}(2022)\citenamefont {Salas-Chavira}, \citenamefont {Fan},\ and\ \citenamefont {Lin}}]{Salas-Chavira:2021wui}%
  \BibitemOpen
  \bibfield  {author} {\bibinfo {author} {\bibfnamefont {Alejandro}\ \bibnamefont {Salas-Chavira}}, \bibinfo {author} {\bibfnamefont {Zhouyou}\ \bibnamefont {Fan}}, \ and\ \bibinfo {author} {\bibfnamefont {Huey-Wen}\ \bibnamefont {Lin}},\ }\bibfield  {title} {\enquote {\bibinfo {title} {{First glimpse into the kaon gluon parton distribution using lattice QCD}},}\ }\href {\doibase 10.1103/PhysRevD.106.094510} {\bibfield  {journal} {\bibinfo  {journal} {Phys. Rev. D}\ }\textbf {\bibinfo {volume} {106}},\ \bibinfo {pages} {094510} (\bibinfo {year} {2022})},\ \Eprint {http://arxiv.org/abs/2112.03124} {arXiv:2112.03124 [hep-lat]} \BibitemShut {NoStop}%
\bibitem [{\citenamefont {Sufian}\ \emph {et~al.}(2019)\citenamefont {Sufian}, \citenamefont {Karpie}, \citenamefont {Egerer}, \citenamefont {Orginos}, \citenamefont {Qiu},\ and\ \citenamefont {Richards}}]{Sufian:2019bol}%
  \BibitemOpen
  \bibfield  {author} {\bibinfo {author} {\bibfnamefont {Raza~Sabbir}\ \bibnamefont {Sufian}}, \bibinfo {author} {\bibfnamefont {Joseph}\ \bibnamefont {Karpie}}, \bibinfo {author} {\bibfnamefont {Colin}\ \bibnamefont {Egerer}}, \bibinfo {author} {\bibfnamefont {Kostas}\ \bibnamefont {Orginos}}, \bibinfo {author} {\bibfnamefont {Jian-Wei}\ \bibnamefont {Qiu}}, \ and\ \bibinfo {author} {\bibfnamefont {David~G.}\ \bibnamefont {Richards}},\ }\bibfield  {title} {\enquote {\bibinfo {title} {{Pion Valence Quark Distribution from Matrix Element Calculated in Lattice QCD}},}\ }\href {\doibase 10.1103/PhysRevD.99.074507} {\bibfield  {journal} {\bibinfo  {journal} {Phys. Rev. D}\ }\textbf {\bibinfo {volume} {99}},\ \bibinfo {pages} {074507} (\bibinfo {year} {2019})},\ \Eprint {http://arxiv.org/abs/1901.03921} {arXiv:1901.03921 [hep-lat]} \BibitemShut {NoStop}%
\bibitem [{\citenamefont {Sufian}\ \emph {et~al.}(2020)\citenamefont {Sufian}, \citenamefont {Egerer}, \citenamefont {Karpie}, \citenamefont {Edwards}, \citenamefont {Jo\'o}, \citenamefont {Ma}, \citenamefont {Orginos}, \citenamefont {Qiu},\ and\ \citenamefont {Richards}}]{Sufian:2020vzb}%
  \BibitemOpen
  \bibfield  {author} {\bibinfo {author} {\bibfnamefont {Raza~Sabbir}\ \bibnamefont {Sufian}}, \bibinfo {author} {\bibfnamefont {Colin}\ \bibnamefont {Egerer}}, \bibinfo {author} {\bibfnamefont {Joseph}\ \bibnamefont {Karpie}}, \bibinfo {author} {\bibfnamefont {Robert~G.}\ \bibnamefont {Edwards}}, \bibinfo {author} {\bibfnamefont {B\'alint}\ \bibnamefont {Jo\'o}}, \bibinfo {author} {\bibfnamefont {Yan-Qing}\ \bibnamefont {Ma}}, \bibinfo {author} {\bibfnamefont {Kostas}\ \bibnamefont {Orginos}}, \bibinfo {author} {\bibfnamefont {Jian-Wei}\ \bibnamefont {Qiu}}, \ and\ \bibinfo {author} {\bibfnamefont {David~G.}\ \bibnamefont {Richards}},\ }\bibfield  {title} {\enquote {\bibinfo {title} {{Pion Valence Quark Distribution from Current-Current Correlation in Lattice QCD}},}\ }\href {\doibase 10.1103/PhysRevD.102.054508} {\bibfield  {journal} {\bibinfo  {journal} {Phys. Rev. D}\ }\textbf {\bibinfo {volume} {102}},\ \bibinfo {pages} {054508} (\bibinfo {year} {2020})},\ \Eprint {http://arxiv.org/abs/2001.04960}
  {arXiv:2001.04960 [hep-lat]} \BibitemShut {NoStop}%
\bibitem [{\citenamefont {Alexandrou}\ \emph {et~al.}(2018)\citenamefont {Alexandrou} \emph {et~al.}}]{Alexandrou:2018egz}%
  \BibitemOpen
  \bibfield  {author} {\bibinfo {author} {\bibfnamefont {Constantia}\ \bibnamefont {Alexandrou}} \emph {et~al.},\ }\bibfield  {title} {\enquote {\bibinfo {title} {{Simulating twisted mass fermions at physical light, strange and charm quark masses}},}\ }\href {\doibase 10.1103/PhysRevD.98.054518} {\bibfield  {journal} {\bibinfo  {journal} {Phys. Rev. D}\ }\textbf {\bibinfo {volume} {98}},\ \bibinfo {pages} {054518} (\bibinfo {year} {2018})},\ \Eprint {http://arxiv.org/abs/1807.00495} {arXiv:1807.00495 [hep-lat]} \BibitemShut {NoStop}%
\bibitem [{\citenamefont {Jay}\ and\ \citenamefont {Neil}(2021)}]{Jay:2020jkz}%
  \BibitemOpen
  \bibfield  {author} {\bibinfo {author} {\bibfnamefont {William~I.}\ \bibnamefont {Jay}}\ and\ \bibinfo {author} {\bibfnamefont {Ethan~T.}\ \bibnamefont {Neil}},\ }\bibfield  {title} {\enquote {\bibinfo {title} {{Bayesian model averaging for analysis of lattice field theory results}},}\ }\href {\doibase 10.1103/PhysRevD.103.114502} {\bibfield  {journal} {\bibinfo  {journal} {Phys. Rev. D}\ }\textbf {\bibinfo {volume} {103}},\ \bibinfo {pages} {114502} (\bibinfo {year} {2021})},\ \Eprint {http://arxiv.org/abs/2008.01069} {arXiv:2008.01069 [stat.ME]} \BibitemShut {NoStop}%
\bibitem [{\citenamefont {Morningstar}\ and\ \citenamefont {Peardon}(2004)}]{Morningstar:2003gk}%
  \BibitemOpen
  \bibfield  {author} {\bibinfo {author} {\bibfnamefont {Colin}\ \bibnamefont {Morningstar}}\ and\ \bibinfo {author} {\bibfnamefont {Mike~J.}\ \bibnamefont {Peardon}},\ }\bibfield  {title} {\enquote {\bibinfo {title} {{Analytic smearing of SU(3) link variables in lattice QCD}},}\ }\href {\doibase 10.1103/PhysRevD.69.054501} {\bibfield  {journal} {\bibinfo  {journal} {Phys. Rev. D}\ }\textbf {\bibinfo {volume} {69}},\ \bibinfo {pages} {054501} (\bibinfo {year} {2004})},\ \Eprint {http://arxiv.org/abs/hep-lat/0311018} {arXiv:hep-lat/0311018} \BibitemShut {NoStop}%
\bibitem [{\citenamefont {{J\"{u}lich Supercomputing Centre}}(2021)}]{JUWELS}%
  \BibitemOpen
  \bibfield  {author} {\bibinfo {author} {\bibnamefont {{J\"{u}lich Supercomputing Centre}}},\ }\bibfield  {title} {\enquote {\bibinfo {title} {{JUWELS Cluster and Booster: Exascale Pathfinder with Modular Supercomputing Architecture at Juelich Supercomputing Centre}},}\ }\href {\doibase 10.17815/jlsrf-7-183} {\bibfield  {journal} {\bibinfo  {journal} {Journal of large-scale research facilities}\ }\textbf {\bibinfo {volume} {7}} (\bibinfo {year} {2021}),\ 10.17815/jlsrf-7-183}\BibitemShut {NoStop}%
\bibitem [{\citenamefont {Jansen}\ and\ \citenamefont {Urbach}(2009)}]{Jansen:2009xp}%
  \BibitemOpen
  \bibfield  {author} {\bibinfo {author} {\bibfnamefont {K.}~\bibnamefont {Jansen}}\ and\ \bibinfo {author} {\bibfnamefont {C.}~\bibnamefont {Urbach}},\ }\bibfield  {title} {\enquote {\bibinfo {title} {{tmLQCD: A Program suite to simulate Wilson Twisted mass Lattice QCD}},}\ }\href {\doibase 10.1016/j.cpc.2009.05.016} {\bibfield  {journal} {\bibinfo  {journal} {Comput.Phys.Commun.}\ }\textbf {\bibinfo {volume} {180}},\ \bibinfo {pages} {2717--2738} (\bibinfo {year} {2009})},\ \Eprint {http://arxiv.org/abs/0905.3331} {arXiv:0905.3331 [hep-lat]} \BibitemShut {NoStop}%
\bibitem [{\citenamefont {Abdel-Rehim}\ \emph {et~al.}(2014)\citenamefont {Abdel-Rehim}, \citenamefont {Burger}, \citenamefont {Deuzeman}, \citenamefont {Jansen}, \citenamefont {Kostrzewa}, \citenamefont {Scorzato},\ and\ \citenamefont {Urbach}}]{Abdel-Rehim:2013wba}%
  \BibitemOpen
  \bibfield  {author} {\bibinfo {author} {\bibfnamefont {Abdou}\ \bibnamefont {Abdel-Rehim}}, \bibinfo {author} {\bibfnamefont {Florian}\ \bibnamefont {Burger}}, \bibinfo {author} {\bibfnamefont {Alber}\ \bibnamefont {Deuzeman}}, \bibinfo {author} {\bibfnamefont {Karl}\ \bibnamefont {Jansen}}, \bibinfo {author} {\bibfnamefont {Bartosz}\ \bibnamefont {Kostrzewa}}, \bibinfo {author} {\bibfnamefont {Luigi}\ \bibnamefont {Scorzato}}, \ and\ \bibinfo {author} {\bibfnamefont {Carsten}\ \bibnamefont {Urbach}},\ }\bibfield  {title} {\enquote {\bibinfo {title} {{Recent developments in the tmLQCD software suite}},}\ }\href {\doibase 10.22323/1.187.0414} {\bibfield  {journal} {\bibinfo  {journal} {PoS}\ }\textbf {\bibinfo {volume} {LATTICE2013}},\ \bibinfo {pages} {414} (\bibinfo {year} {2014})},\ \Eprint {http://arxiv.org/abs/1311.5495} {arXiv:1311.5495 [hep-lat]} \BibitemShut {NoStop}%
\bibitem [{\citenamefont {Deuzeman}\ \emph {et~al.}(2013)\citenamefont {Deuzeman}, \citenamefont {Jansen}, \citenamefont {Kostrzewa},\ and\ \citenamefont {Urbach}}]{Deuzeman:2013xaa}%
  \BibitemOpen
  \bibfield  {author} {\bibinfo {author} {\bibfnamefont {A.}~\bibnamefont {Deuzeman}}, \bibinfo {author} {\bibfnamefont {K.}~\bibnamefont {Jansen}}, \bibinfo {author} {\bibfnamefont {B.}~\bibnamefont {Kostrzewa}}, \ and\ \bibinfo {author} {\bibfnamefont {C.}~\bibnamefont {Urbach}},\ }\bibfield  {title} {\enquote {\bibinfo {title} {{Experiences with OpenMP in tmLQCD}},}\ }\href@noop {} {\bibfield  {journal} {\bibinfo  {journal} {PoS}\ }\textbf {\bibinfo {volume} {LATTICE2013}},\ \bibinfo {pages} {416} (\bibinfo {year} {2013})},\ \Eprint {http://arxiv.org/abs/1311.4521} {arXiv:1311.4521 [hep-lat]} \BibitemShut {NoStop}%
\bibitem [{\citenamefont {Deuzeman}\ \emph {et~al.}(2012)\citenamefont {Deuzeman}, \citenamefont {Reker},\ and\ \citenamefont {Urbach}}]{Deuzeman:2011wz}%
  \BibitemOpen
  \bibfield  {author} {\bibinfo {author} {\bibfnamefont {Albert}\ \bibnamefont {Deuzeman}}, \bibinfo {author} {\bibfnamefont {Siebren}\ \bibnamefont {Reker}}, \ and\ \bibinfo {author} {\bibfnamefont {Carsten}\ \bibnamefont {Urbach}} (\bibinfo {collaboration} {ETM}),\ }\bibfield  {title} {\enquote {\bibinfo {title} {{Lemon: an MPI parallel I/O library for data encapsulation using LIME}},}\ }\href {\doibase 10.1016/j.cpc.2012.01.016} {\bibfield  {journal} {\bibinfo  {journal} {Comput. Phys. Commun.}\ }\textbf {\bibinfo {volume} {183}},\ \bibinfo {pages} {1321--1335} (\bibinfo {year} {2012})},\ \Eprint {http://arxiv.org/abs/1106.4177} {arXiv:1106.4177 [hep-lat]} \BibitemShut {NoStop}%
\bibitem [{\citenamefont {Clark}\ \emph {et~al.}(2010)\citenamefont {Clark}, \citenamefont {Babich}, \citenamefont {Barros}, \citenamefont {Brower},\ and\ \citenamefont {Rebbi}}]{Clark:2009wm}%
  \BibitemOpen
  \bibfield  {author} {\bibinfo {author} {\bibfnamefont {M.~A.}\ \bibnamefont {Clark}}, \bibinfo {author} {\bibfnamefont {R.}~\bibnamefont {Babich}}, \bibinfo {author} {\bibfnamefont {K.}~\bibnamefont {Barros}}, \bibinfo {author} {\bibfnamefont {R.~C.}\ \bibnamefont {Brower}}, \ and\ \bibinfo {author} {\bibfnamefont {C.}~\bibnamefont {Rebbi}},\ }\bibfield  {title} {\enquote {\bibinfo {title} {{Solving Lattice QCD systems of equations using mixed precision solvers on GPUs}},}\ }\href {\doibase 10.1016/j.cpc.2010.05.002} {\bibfield  {journal} {\bibinfo  {journal} {Comput. Phys. Commun.}\ }\textbf {\bibinfo {volume} {181}},\ \bibinfo {pages} {1517--1528} (\bibinfo {year} {2010})},\ \Eprint {http://arxiv.org/abs/0911.3191} {arXiv:0911.3191 [hep-lat]} \BibitemShut {NoStop}%
\bibitem [{\citenamefont {Babich}\ \emph {et~al.}(2011)\citenamefont {Babich}, \citenamefont {Clark}, \citenamefont {Joo}, \citenamefont {Shi}, \citenamefont {Brower},\ and\ \citenamefont {Gottlieb}}]{Babich:2011np}%
  \BibitemOpen
  \bibfield  {author} {\bibinfo {author} {\bibfnamefont {R.}~\bibnamefont {Babich}}, \bibinfo {author} {\bibfnamefont {M.~A.}\ \bibnamefont {Clark}}, \bibinfo {author} {\bibfnamefont {B.}~\bibnamefont {Joo}}, \bibinfo {author} {\bibfnamefont {G.}~\bibnamefont {Shi}}, \bibinfo {author} {\bibfnamefont {R.~C.}\ \bibnamefont {Brower}}, \ and\ \bibinfo {author} {\bibfnamefont {S.}~\bibnamefont {Gottlieb}},\ }\bibfield  {title} {\enquote {\bibinfo {title} {{Scaling Lattice QCD beyond 100 GPUs}},}\ }in\ \href {\doibase 10.1145/2063384.2063478} {\emph {\bibinfo {booktitle} {{SC11 International Conference for High Performance Computing, Networking, Storage and Analysis Seattle, Washington, November 12-18, 2011}}}}\ (\bibinfo {year} {2011})\ \Eprint {http://arxiv.org/abs/1109.2935} {arXiv:1109.2935 [hep-lat]} \BibitemShut {NoStop}%
\bibitem [{\citenamefont {Clark}\ \emph {et~al.}(2016)\citenamefont {Clark}, \citenamefont {Joó}, \citenamefont {Strelchenko}, \citenamefont {Cheng}, \citenamefont {Gambhir},\ and\ \citenamefont {Brower}}]{Clark:2016rdz}%
  \BibitemOpen
  \bibfield  {author} {\bibinfo {author} {\bibfnamefont {M.~A.}\ \bibnamefont {Clark}}, \bibinfo {author} {\bibfnamefont {Bálint}\ \bibnamefont {Joó}}, \bibinfo {author} {\bibfnamefont {Alexei}\ \bibnamefont {Strelchenko}}, \bibinfo {author} {\bibfnamefont {Michael}\ \bibnamefont {Cheng}}, \bibinfo {author} {\bibfnamefont {Arjun}\ \bibnamefont {Gambhir}}, \ and\ \bibinfo {author} {\bibfnamefont {Richard}\ \bibnamefont {Brower}},\ }\bibfield  {title} {\enquote {\bibinfo {title} {{Accelerating Lattice QCD Multigrid on GPUs Using Fine-Grained Parallelization}},}\ }\href@noop {} {\  (\bibinfo {year} {2016})},\ \Eprint {http://arxiv.org/abs/1612.07873} {arXiv:1612.07873 [hep-lat]} \BibitemShut {NoStop}%
\bibitem [{\citenamefont {{R Core Team}}(2019)}]{R:2019}%
  \BibitemOpen
  \bibfield  {author} {\bibinfo {author} {\bibnamefont {{R Core Team}}},\ }\href {https://www.R-project.org/} {\emph {\bibinfo {title} {R: A Language and Environment for Statistical Computing}}},\ \bibinfo {organization} {R Foundation for Statistical Computing},\ \bibinfo {address} {Vienna, Austria} (\bibinfo {year} {2019})\BibitemShut {NoStop}%
\bibitem [{\citenamefont {Petschlies~et al.}(2024)}]{CVC:2024}%
  \BibitemOpen
  \bibfield  {author} {\bibinfo {author} {\bibfnamefont {M.}~\bibnamefont {Petschlies~et al.}},\ }\href@noop {} {\enquote {\bibinfo {title} {cvc package},}\ }\bibinfo {howpublished} {\url{https://github.com/marcuspetschlies/cvc}} (\bibinfo {year} {2024})\BibitemShut {NoStop}%
\bibitem [{\citenamefont {Alexandrou}\ \emph {et~al.}(2017)\citenamefont {Alexandrou}, \citenamefont {Constantinou},\ and\ \citenamefont {Panagopoulos}}]{Alexandrou:2015sea}%
  \BibitemOpen
  \bibfield  {author} {\bibinfo {author} {\bibfnamefont {Constantia}\ \bibnamefont {Alexandrou}}, \bibinfo {author} {\bibfnamefont {Martha}\ \bibnamefont {Constantinou}}, \ and\ \bibinfo {author} {\bibfnamefont {Haralambos}\ \bibnamefont {Panagopoulos}} (\bibinfo {collaboration} {ETM}),\ }\bibfield  {title} {\enquote {\bibinfo {title} {{Renormalization functions for Nf=2 and Nf=4 twisted mass fermions}},}\ }\href {\doibase 10.1103/PhysRevD.95.034505} {\bibfield  {journal} {\bibinfo  {journal} {Phys. Rev. D}\ }\textbf {\bibinfo {volume} {95}},\ \bibinfo {pages} {034505} (\bibinfo {year} {2017})},\ \Eprint {http://arxiv.org/abs/1509.00213} {arXiv:1509.00213 [hep-lat]} \BibitemShut {NoStop}%
\bibitem [{\citenamefont {Caracciolo}\ \emph {et~al.}(1992)\citenamefont {Caracciolo}, \citenamefont {Menotti},\ and\ \citenamefont {Pelissetto}}]{Caracciolo:1991cp}%
  \BibitemOpen
  \bibfield  {author} {\bibinfo {author} {\bibfnamefont {Sergio}\ \bibnamefont {Caracciolo}}, \bibinfo {author} {\bibfnamefont {Pietro}\ \bibnamefont {Menotti}}, \ and\ \bibinfo {author} {\bibfnamefont {Andrea}\ \bibnamefont {Pelissetto}},\ }\bibfield  {title} {\enquote {\bibinfo {title} {{One loop analytic computation of the energy momentum tensor for lattice gauge theories}},}\ }\href {\doibase 10.1016/0550-3213(92)90339-D} {\bibfield  {journal} {\bibinfo  {journal} {Nucl. Phys. B}\ }\textbf {\bibinfo {volume} {375}},\ \bibinfo {pages} {195--239} (\bibinfo {year} {1992})}\BibitemShut {NoStop}%
\bibitem [{\citenamefont {Panagopoulos}\ \emph {et~al.}(2021)\citenamefont {Panagopoulos}, \citenamefont {Panagopoulos},\ and\ \citenamefont {Spanoudes}}]{Panagopoulos:2020qcn}%
  \BibitemOpen
  \bibfield  {author} {\bibinfo {author} {\bibfnamefont {George}\ \bibnamefont {Panagopoulos}}, \bibinfo {author} {\bibfnamefont {Haralambos}\ \bibnamefont {Panagopoulos}}, \ and\ \bibinfo {author} {\bibfnamefont {Gregoris}\ \bibnamefont {Spanoudes}},\ }\bibfield  {title} {\enquote {\bibinfo {title} {{Two-loop renormalization and mixing of gluon and quark energy-momentum tensor operators}},}\ }\href {\doibase 10.1103/PhysRevD.103.014515} {\bibfield  {journal} {\bibinfo  {journal} {Phys. Rev. D}\ }\textbf {\bibinfo {volume} {103}},\ \bibinfo {pages} {014515} (\bibinfo {year} {2021})},\ \Eprint {http://arxiv.org/abs/2010.02062} {arXiv:2010.02062 [hep-lat]} \BibitemShut {NoStop}%
\bibitem [{\citenamefont {Alexandrou}\ \emph {et~al.}(2020)\citenamefont {Alexandrou}, \citenamefont {Athenodorou}, \citenamefont {Cichy}, \citenamefont {Dromard}, \citenamefont {Garcia-Ramos}, \citenamefont {Jansen}, \citenamefont {Wenger},\ and\ \citenamefont {Zimmermann}}]{Alexandrou:2017hqw}%
  \BibitemOpen
  \bibfield  {author} {\bibinfo {author} {\bibfnamefont {Constantia}\ \bibnamefont {Alexandrou}}, \bibinfo {author} {\bibfnamefont {Andreas}\ \bibnamefont {Athenodorou}}, \bibinfo {author} {\bibfnamefont {Krzysztof}\ \bibnamefont {Cichy}}, \bibinfo {author} {\bibfnamefont {Arthur}\ \bibnamefont {Dromard}}, \bibinfo {author} {\bibfnamefont {Elena}\ \bibnamefont {Garcia-Ramos}}, \bibinfo {author} {\bibfnamefont {Karl}\ \bibnamefont {Jansen}}, \bibinfo {author} {\bibfnamefont {Urs}\ \bibnamefont {Wenger}}, \ and\ \bibinfo {author} {\bibfnamefont {Falk}\ \bibnamefont {Zimmermann}},\ }\bibfield  {title} {\enquote {\bibinfo {title} {{Comparison of topological charge definitions in Lattice QCD}},}\ }\href {\doibase 10.1140/epjc/s10052-020-7984-9} {\bibfield  {journal} {\bibinfo  {journal} {Eur. Phys. J. C}\ }\textbf {\bibinfo {volume} {80}},\ \bibinfo {pages} {424} (\bibinfo {year} {2020})},\ \Eprint {http://arxiv.org/abs/1708.00696} {arXiv:1708.00696 [hep-lat]} \BibitemShut {NoStop}%
\end{thebibliography}%

\pagebreak

\section{Supplemental Material}
\label{sec:supplementaries}

\subsection{Results for the bare and renormalized momentum fractions}

\begin{figure*}[h]
    \centering
    \includegraphics[width = \textwidth]{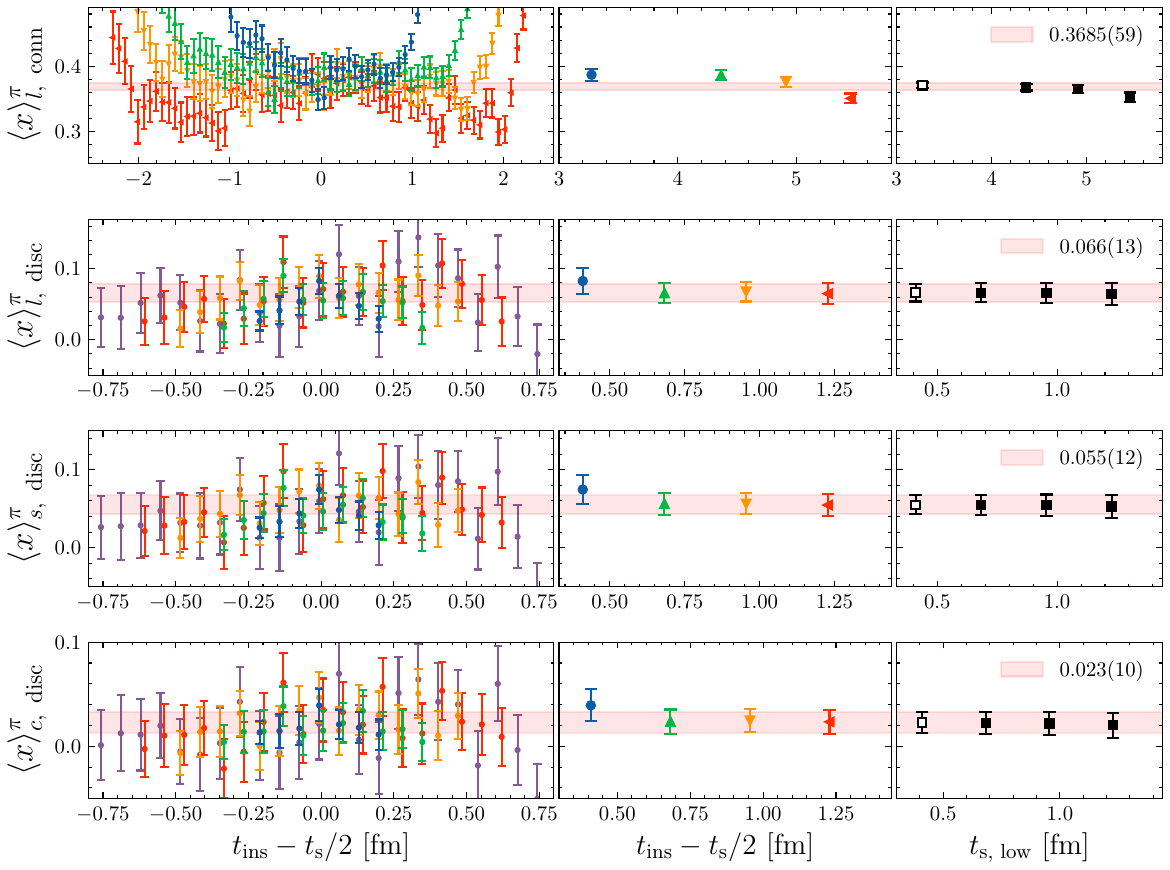}
    \caption{Complete decomposition of the bare momentum fraction carried by quarks for the pion using  the C ensemble. The two upper rows give the momentum fraction $\langle x \rangle_q $ carried by $u+d$  (top for connected and second row from top for disconnected), the 3rd row   for the strange quark and the lower row for the charm quark. In the first column we show the bare ratio for different values of $t_s$, in the second column the individual fits per $t_s$, and in the last column the plateau fit as a function of the lowest value of $t_s$, $t_{s,\textrm{low}}$, used in the fits  which demonstrates the convergence as a function of  $t_{s,\textrm{low}}$ varies. The band shows the value after model averaging and the open symbol on the right panel corresponds to the fit with the highest probability. 
    }
    \label{fig:plateau_cB64_pion_bare}
\end{figure*}

\begin{figure*}[h]
    \centering
    \includegraphics[width = \textwidth]{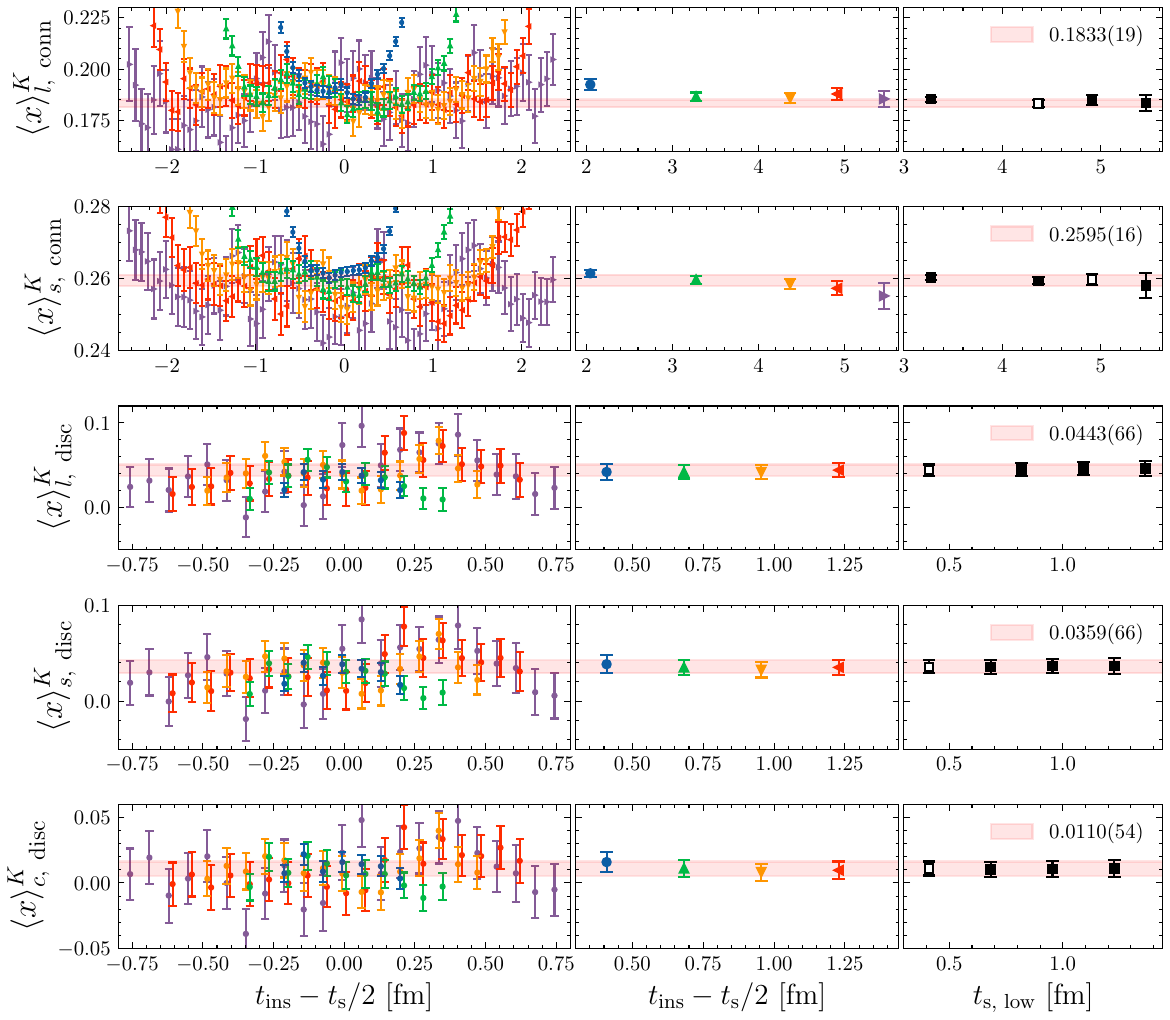}
    \caption{Complete decomposition of the momentum carried by quarks in  the kaon using the C ensemble. The notation is the same as that of  Fig.\ref{fig:plateau_cB64_pion_bare}. }
    \label{fig:plateau_cB64_kaon_bare}
\end{figure*}

The bare momentum fractions can be extracted from 
the ratios in Eq.~(\ref{eq:Rdef}) by performing a 
simultaneous fit for all $\tsnk > t_{s,\textrm{low}}$. 
In this way, from the set of all available $t_s$,
$t_{s,\textrm{low}}$ marks the lowest value within this set from
where we choose to fit.
Figs.~\ref{fig:plateau_cB64_pion_bare} 
and~\ref{fig:plateau_cB64_kaon_bare} show the
overlay of the raw ratio data for individual source-sink
separations $\tsnk$ in the left-hand column and
the fit for individual values of $\tsnk$ in the center
for the pion and the kaon, respectively,
using the B ensemble as defined in Table~\ref{tab:ensembles}. 
Finally, the right-hand column shows the stability of the fit as we vary the
lower bound $t_{s,\textrm{low}}$ of data sets included in the
simultaneous fit.
In ~Tables~\ref{tab:bareresults-cB64}\nobreakdash--\ref{tab:bareresults-cD96} 
we compile all the bare quantities extracted 
using this procedure for each ensemble. 
We show connected contributions as well as
disconnected contributions and the component of 
the EMT used to extract those values for 
both the pion and the kaon.

\begin{figure}
    \centering
    \includegraphics[width = 0.47 \textwidth]{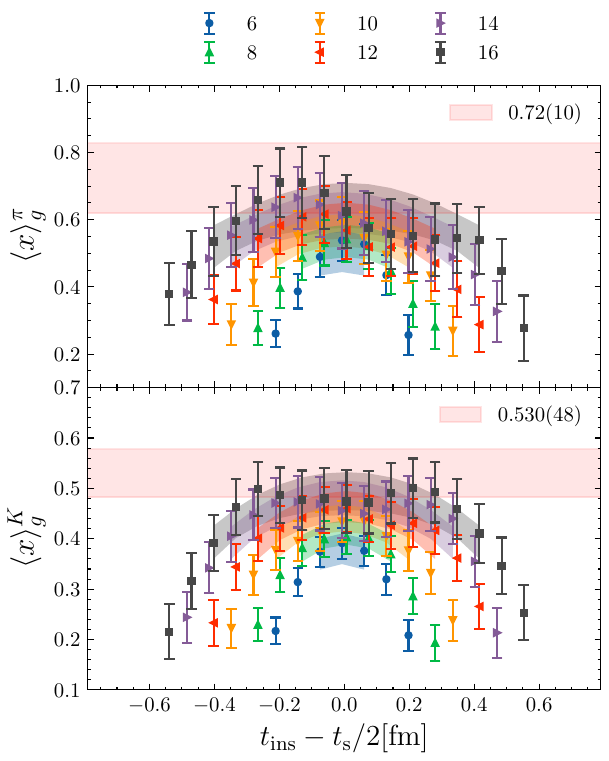}
    \caption{Two state fit for both pion and kaon at 10 stout-smearing steps. The bands in colors are produced with the values of the fits to the correlators. The horizontal band shows the value after model averaging.}
    \label{fig:two-state}
\end{figure}

\begin{figure}
    \centering
    \includegraphics[width = 0.47 \textwidth]{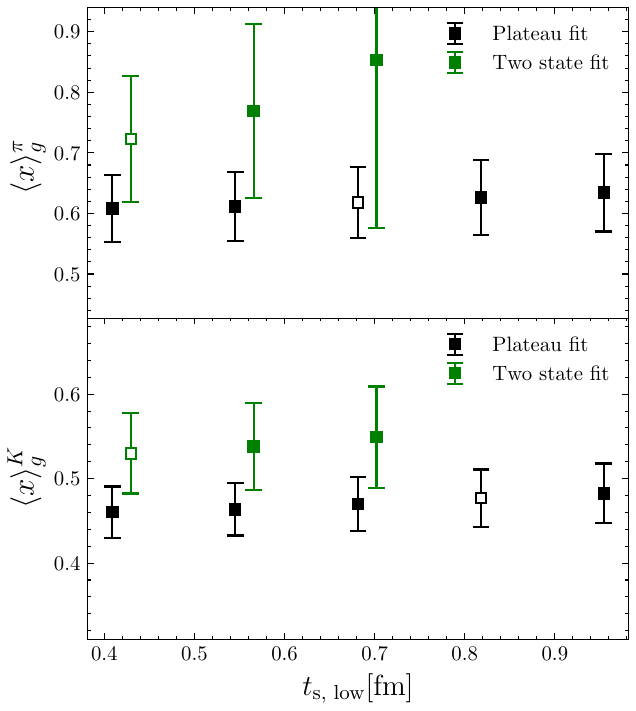}
    \caption{Plateau and two-state fit for $\langle x \rangle_g$ for the pion(upper panel) and the kaon(lower panel). The black squares represent the values per $t_{s, low}$ for the plateau fit and the green squares the values for the two-state fit. Both fits agree, while the plateau shows a good stability at higher values of $t_{s, low}$ the two-state fit gains error faster. The open symbol shows the values with the highest probability for each fit.}
    \label{fig:two-state_comp}
\end{figure}

In the case of the gluon contribution, we can perform two-state fits to the 
two- and three-point functions and compare them with the plateau fit results. 
In Fig.~\ref{fig:two-state} we show the curves for the two-state fit and 
in Fig.~\ref{fig:two-state_comp} we show the comparison between the values 
extracted from the two-state fit and the plateau fit, and they
are in agreement within 
errors. Because the errors increase very fast for the two-state fit,
we decided to use only the plateau fit results in our analysis.


The renormalized results for the three ensembles 
are presented in Tables~\ref{tab:res_cB64}, ~\ref{tab:res_cC80},
and~\ref{tab:res_cD96}. The nominal difference of the renormalized results
for the pion with our previous analysis
on ensemble B (cB211.072.64) in Ref.~\cite{ExtendedTwistedMass:2021rdx}
is rooted in the improvement of available statistics, especially for
the gluon operator insertion, as well as the improvement in the determination of the
renormalization and mixing coefficients.

\begin{figure}[h]
    \centering
    \includegraphics{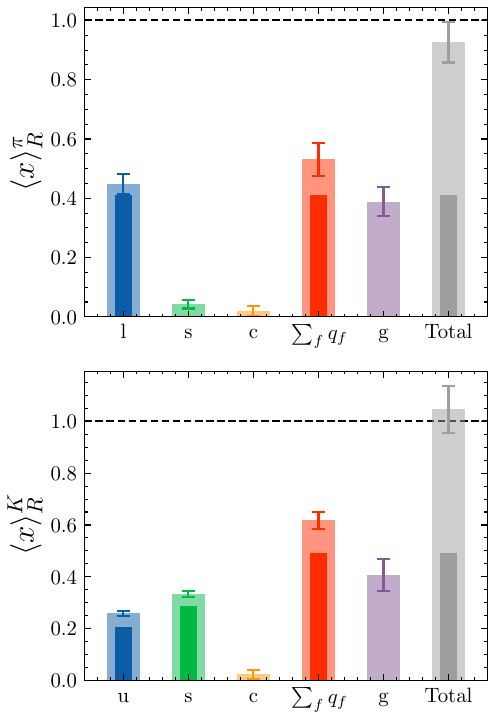}
    \caption{The quark and gluon momentum fractions for the pion (upper panel) and kaon (lower panel) using the numbers in Table~\ref{tab:results_ours}. Inner bars represent only the connected contributions, while the outer bars show the total,  including disconnected contributions.}
    \label{fig:bars}
\end{figure}

In Fig.~\ref{fig:bars} we show the individual quark 
and gluon momentum fractions in the pion and kaon at 
the continuum limit using the numbers of Table~\ref{tab:results_ours}.
The continuum limit has been calculated by fitting the 
results for each ensemble to a constant (including the 
three points and also the two points closer to the continuum) 
and the linear fit in $a^2$. In Table \ref{tab:weights} we 
present the result of these fits and the probabilities associated to each one.
%


\begin{table}[th]
\caption{Weights and values for each fit contributing to the continuum limit. Each row represents a different contribution to the momentum, from quarks or gluons, and then their sum for the pion and for the kaon. Each column represents the weight and value for each fit taken, with the first one the constant fit taking only two points, the second one the constant fit with three points and the last one the linear fit taking all three points.}
  \centering
  \begin{tabular*}{.49\textwidth}{@{\extracolsep{\fill}}cccl}
    \toprule\hline
    & Plateau 2 points & Plateau 3 points & Linear fit \\
    \midrule\hline
    \(\langle x\rangle^{\pi}_{q, R}\) & $0.218, 0.539(30)$ & $0.523, 0.552(15)$ & $0.259, 0.486(88)$\\
    \(\langle x\rangle^{\pi}_{g, R}\) & $0.304, 0.403(35)$ & $0.504, 0.375(21)$ & $0.191, 0.398(94)$\\
    \(\langle x\rangle^{\pi}_{R}\) & $0.237, 0.948(47)$ & $0.555, 0.926(26)$ & $0.208, 0.90(13)$\\
    \(\langle x\rangle^{K}_{q, R}\) & $0.208, 0.609(18)$ & $0.554, 0.612(10)$  & $0.238, 0.642(55)$\\
    \(\langle x\rangle^{K}_{g, R}\) & $0.317, 0.386(22)$ & $0.296, 0.361(13)$ & $0.387, 0.461(64)$\\
    \(\langle x \rangle^{K}_{R}\) & $0.211, 0.998(28)$ & $0.310, 0.973(17)$ & $0.478, 1.115(85)$\\
    \bottomrule\hline
  \end{tabular*}
  \label{tab:weights}
\end{table}

\label{ssec:bare_results}


\begin{table}[th]
\caption{Bare values for the different contributions for the B (cB211.072.64) ensemble.}
  \centering
  \begin{tabular*}{.49\textwidth}{@{\extracolsep{\fill}}lccc}
    \toprule\hline
    Contribution & Operator & $\pi $& $K$ \\
    \midrule\hline
    \(\langle x\rangle_l^\mathrm{conn}\) & \(\bar{T}_{44}\) & 0.3929(44) & 0.1926(13) \\
    \(\langle x\rangle_s^\mathrm{conn}\) & \(\bar{T}_{44}\) & - - -      & 0.2694(06) \\
    \(\langle x\rangle_l^\mathrm{disc}\) & \(\bar{T}_{4k}\) & 0.0612(59) & 0.0510(34) \\
    \(\langle x\rangle_s^\mathrm{disc}\) & \(\bar{T}_{4k}\) & 0.0441(58) & 0.0385(32) \\
    \(\langle x\rangle_c^\mathrm{disc}\) & \(\bar{T}_{4k}\) & 0.0150(46) & 0.0137(23) \\
  \bottomrule\hline
  \end{tabular*}
  \label{tab:bareresults-cB64}
\end{table}

\begin{table}[th]
\caption{Bare values for the different contributions to $\langle x \rangle$ for the  C (cC211.060.80) ensemble.}
  \centering
  \begin{tabular*}{.49\textwidth}{@{\extracolsep{\fill}}lccc}
    \toprule\hline
    Contribution & Operator & $\pi $& $K$ \\
    \midrule\hline
    \(\langle x\rangle_l^\mathrm{conn}\) & \(\bar{T}_{44}\) & 0.369(06) & 0.183(02) \\
    \(\langle x\rangle_s^\mathrm{conn}\) & \(\bar{T}_{44}\) & - - - & 0.260(02) \\
    \(\langle x\rangle_l^\mathrm{disc}\) & \(\bar{T}_{4k}\) & 0.066(13) & 0.044(07) \\
    \(\langle x\rangle_s^\mathrm{disc}\) & \(\bar{T}_{4k}\) & 0.055(12) & 0.036(07) \\
    \(\langle x\rangle_c^\mathrm{disc}\) & \(\bar{T}_{4k}\) & 0.023(10) & 0.011(05) \\
  \bottomrule\hline
  \end{tabular*}

  \label{tab:bareresults-cC80}
\end{table}

\begin{table}[!th]
\caption{Bare values for the different contributions to $\langle x \rangle$ for the D (cD211.054.96) ensemble.}
  \centering
  \begin{tabular*}{.49\textwidth}{@{\extracolsep{\fill}}lccc}
    \toprule\hline
    Contribution & Operator & $\pi $& $K$ \\
    \midrule\hline
    \(\langle x\rangle_l^\mathrm{conn}\) & \(\bar{T}_{44}\) & 0.357(03) & 0.180(02) \\
    \(\langle x\rangle_s^\mathrm{conn}\) & \(\bar{T}_{44}\) & - - - & 0.249(01) \\
    \(\langle x\rangle_l^\mathrm{disc}\) & \(\bar{T}_{4k}\) & 0.041(11) & 0.055(07) \\
    \(\langle x\rangle_s^\mathrm{disc}\) & \(\bar{T}_{4k}\) & 0.029(11) & 0.047(07) \\
    \(\langle x\rangle_c^\mathrm{disc}\) & \(\bar{T}_{4k}\) & 0.010(19) & 0.045(11) \\
  \bottomrule\hline
  \end{tabular*}
  \label{tab:bareresults-cD96}
\end{table}

\begin{table}[th]
\caption{Renormalized values for the different contributions for B (cB211.072.64) ensemble in the $\MSbar$ scheme at a scale of 2~GeV.}
  \centering
  \begin{tabular*}{.49\textwidth}{@{\extracolsep{\fill}}lcc}
    \toprule\hline
    Contribution  & $\pi $& $K$ \\
    \midrule\hline
    \(\langle x\rangle_{l,R}\)   & $0.495(09)$ & $0.264(05)$ \\
    \(\langle x\rangle_{s,R}\)   & $0.047(07)$ & $0.337(07)$ \\
    \(\langle x\rangle_{c,R}\)   & $0.014(05)$ & $0.013(03)$ \\
    \(\langle x\rangle_{q,R}\)   & $0.557(18)$ & $0.613(13)$ \\
    \(\langle x\rangle_{g,R}\)   & $0.360(25)$ & $0.346(16)$ \\
    
  \bottomrule\hline
  \end{tabular*}
  \label{tab:res_cB64}
\end{table}

\begin{table}[th]
 \caption{Renormalized values for the different contributions for C (cC211.060.80) ensemble in the $\MSbar$ scheme at a scale of 2~GeV.}
  \centering
  \begin{tabular*}{.49\textwidth}{@{\extracolsep{\fill}}lcc}
    \toprule\hline
    Contribution  & $\pi $& $K$ \\
    \midrule\hline
    \(\langle x\rangle_{l,R}\)   & $0.479(15)$ & $0.249(08)$ \\
    \(\langle x\rangle_{s,R}\)   & $0.059(14)$ & $0.327(08)$ \\
    \(\langle x\rangle_{c,R}\)   & $0.022(11)$ & $0.009(06)$ \\
    \(\langle x\rangle_{q,R}\)   & $0.560(38)$ & $0.585(21)$ \\
    \(\langle x\rangle_{g,R}\)   & $0.478(51)$ & $0.375(28)$ \\
    
  \bottomrule\hline
  \end{tabular*}
  \label{tab:res_cC80}
\end{table}

\begin{table}[th]
 \caption{Renormalized values for the different contributions for D (cD211.054.96) ensemble in the $\MSbar$ scheme at a scale of 2~GeV.}
  \label{tab:res_cD96}
  \centering
  \begin{tabular*}{.49\textwidth}{@{\extracolsep{\fill}}lcc}
    \toprule\hline
    Contribution  & $\pi $& $K$ \\
    \midrule\hline
    \(\langle x\rangle_{l,R}\)   & $0.459(15)$ & $0.270(09)$ \\
    \(\langle x\rangle_{s,R}\)   & $0.033(13)$ & $ 0.342(12)$ \\
    \(\langle x\rangle_{c,R}\)   & $0.012(22)$ & $ 0.052(13)$ \\
    \(\langle x\rangle_{q,R}\)   & $0.504(50)$ & $0.664(32)$ \\
    \(\langle x\rangle_{g,R}\)   & $0.337(48)$ & $0.404(35)$ \\
  \bottomrule\hline
  \end{tabular*}
\end{table}

\subsection{Renormalization in the $\MSbar$ scheme}
\label{ssec:renormalization}
The bare quark and gluon average momentum fractions have been renormalized nonperturbatively by using the RI$'$/MOM scheme followed by perturbative conversion to $\overline{\rm MS}$ at the scale $\mu = 2$ GeV. Based on global symmetries, the singlet quark and gluon EMTs in Eqs.~(\ref{Tq} -- \ref{Tg}) mix under renormalization according to Eq.~(\ref{eq:mixing}).
%
%
Due to the breaking of rotational symmetry on the lattice, $\bar{T}^{q(g)}_{\mu=\nu}$ and $\bar{T}^{q(g)}_{\mu \neq \nu}$ belong to different representations of the hypercubic group, and they are renormalized differently. The RI$'$/MOM renormalization matrix for each case is calculated by the following conditions:
\begin{equation}
  {\begin{pmatrix}
    Z_{qq}^{s, {\rm RI}'} & Z_{qg}^{{\rm RI}'} \\
    Z_{gq}^{{\rm RI}'} & Z_{gg}^{{\rm RI}'} 
  \end{pmatrix}}^{-1} = 
  \begin{pmatrix}
   \frac{1}{12} {\rm Tr} [V_{qq} (V_{qq}^{\rm tree})^{-1}] & V_{qg} (V_{gg}^{\rm tree})^{-1} \\
   \frac{1}{12} {\rm Tr} [V_{gq} (V_{qq}^{\rm tree})^{-1}] & V_{gg} (V_{gg}^{\rm tree})^{-1}
  \end{pmatrix},
\end{equation}
where
\begin{eqnarray}
 V_{qq} &=& \frac{\langle q(p) \bar{T}^q_{\mu \nu} \bar{q} (p) \rangle_{\rm amp}}{Z_q^{{\rm RI}'}}, \\ 
 V_{qg} &=& \frac{\langle {\rm Tr} [A_\rho (p) A_\rho (-p)] \ \bar{T}^q_{\mu \nu} \rangle}{\langle {\rm Tr} [A_\rho (p) A_\rho (-p)] \rangle}, \\
V_{gq} &=& \frac{\langle q(p) \bar{T}^g_{\mu \nu} \bar{q} (p) \rangle_{\rm amp}}{Z_q^{{\rm RI}'}}, \\
V_{gg} &=& \frac{\langle {\rm Tr} [A_\rho (p) A_\rho (-p)] \ \bar{T}^g_{\mu \nu} \rangle}{\langle {\rm Tr} [A_\rho (p) A_\rho (-p)] \rangle},
\end{eqnarray}
\noindent
and $V_{qq}^{\rm tree}$ ($V_{gg}^{\rm tree}$) is the tree-level value of $V_{qq}$ ($V_{gg}$). The renormalization factor of the quark field $Z_q^{{\rm RI}'}$ is calculated in standard way through the quark propagator~\cite{Alexandrou:2015sea}. Note that in $V_{gg}$ and $V_{qg}$, we set $\rho \neq (\mu,\nu)$ and $p_\rho=0$ in order to cancel the mixing of EMTs with gauge-noninvariant operators~\cite{Caracciolo:1991cp}, at least, to the leading order. For the same reason, in the case $\mu=\nu$, $\bar{T}_{\mu\nu}^{q (g)}$ is replaced by $\bar{T}_{\mu\mu}^{q (g)} - \bar{T}_{\sigma\sigma}^{q (g)}$, $\sigma \neq (\mu,\rho)$. The corresponding $\overline{\rm MS}$ renormalization matrix is extracted through:
\begin{eqnarray}
  {\begin{pmatrix}
    Z_{qq}^{s, \overline{\rm MS}} & Z_{qg}^{\overline{\rm MS}} \nonumber\\
    Z_{gq}^{\overline{\rm MS}} & Z_{gg}^{\overline{\rm MS}} 
  \end{pmatrix}}^{-1} &=& 
  {\begin{pmatrix}
    Z_{qq}^{s, {\rm RI}'} & Z_{qg}^{{\rm RI}'} \\
    Z_{gq}^{{\rm RI}'} & Z_{gg}^{{\rm RI}'} 
  \end{pmatrix}}^{-1} \\
   & \times &
  {\begin{pmatrix}
  C_{qq}^{\overline{\rm MS},{\rm RI}'} & C_{qg}^{\overline{\rm MS},{\rm RI}'} \\
    C_{gq}^{\overline{\rm MS},{\rm RI}'} & C_{gg}^{\overline{\rm MS},{\rm RI}'} 
  \end{pmatrix}}^{-1},
\end{eqnarray}
 where $C_{ij}^{\overline{\rm MS},{\rm RI}'}$ are the conversion-matrix coefficients calculated up to two-loop order~\cite{Panagopoulos:2020qcn}.  For the renormalization of flavor-nonsinglet combinations $\langle x \rangle_{u+d-2s}$ and $\langle x \rangle_{u+d+s-3c}$, the nonsinglet renormalization factor $Z_{qq}$ is used, which is calculated from the connected part of $V_{qq}$ (see ~\cite{Alexandrou:2015sea}).
 
 In this work, the diagonal elements  $Z_{qq}^{s, \overline{\rm MS}}$ and $Z_{gg}^{\overline{\rm MS}}$ are evaluated non-perturbatively, while the off-diagonal $Z_{qg}^{\overline{\rm MS}}$ and $Z_{gq}^{\overline{\rm MS}}$ are calculated in one-loop lattice perturbation theory.  For the non-perturbative calculations, we have employed high-statistical ($\sim 10000 - 30000$ configurations) $N_f=4$ mass-degenerate ensembles generated by ETMC at the same lattice spacings of the three physical ensembles (see Table~\ref{tab:ensembles}). The parameters of the $N_f=4$ ensembles are given in Table~\ref{tab:Nf4ensembles}.

\begin{table}[!h]
\caption{$N_f=4$ ensembles and their parameters used for the renormalization: $\beta = (2 N/g^2)$, lattice volume ($L^3 \times T$), twisted-mass parameter $(a \mu)$, hopping parameter $\kappa$ and clover coefficient $c_{\rm SW}$.}
  \centering
\begin{tabular} {cccccc}
  \hline
  Ensemble & & & & & \\
   Label & $\beta$ & $L^3 \times T$ & $a \mu$ & $\kappa$ & $c_{\rm SW}$ \\
   \hline
   B  & 1.778 & $12^3 \times 24$ & 0.006 & 0.1393050 & 1.6900 \\
   C  & 1.836 & $24^3 \times 48$ & 0.005 & 0.1386735 & 1.6452 \\
   D  & 1.900 & $24^3 \times 48$ & 0.004 & 0.1379346 & 1.6112 \\ 
   \hline
\end{tabular}
    \label{tab:Nf4ensembles}
    \end{table}
 
 An important aspect of our renormalization program is the improvement of the nonperturbative estimates by subtracting tree-level or one-loop discretization errors~\cite{Alexandrou:2015sea}. Stout smearing is employed in the gluon links of $\bar{T}_{\mu\nu}^g$ with the same set of smearing levels as used  during the computation of the matrix elements for ensembles B, C and D. In the perturbative calculation of $Z_{gq}^{\overline{\rm MS}}$, we restrict the stout smearing to three steps to manage the rapid growth of algebraic expressions, which can result in millions of terms when more stout-smearing steps are included. We have also employed Wilson flow (instead of stout smearing) in the nonperturbative calculations and we have confirmed the equivalence of the two smearing methods \cite{Alexandrou:2017hqw} when $\tau = \rstout \, \nstout$ ($\tau$ is the flow time, and $\rstout$, $\nstout$ are the stout parameter and number of stout-smearing steps, respectively). In table \ref{tab:Zfactors}, we collect our results for the renormalization coefficients that enter our analysis. 

\begin{table}[!thb]
 \caption{Renormalization coefficients in the $\overline{\rm MS}$ scheme at 2 GeV for the three ensembles of Table~\ref{tab:Nf4ensembles} and for different tensor structures. For the computation of $Z_{gg}$, 5-10 steps of stout smearing are employed. The error quoted in the parenthesis is statistical; $Z_{qg}$ and $Z_{gq}$ have no statistical error because they are calculated in perturbation theory.}
  \centering
  \begin{tabular}{l|c|c|c}
  \hline 
  \ & \ Ensemble & \ Ensemble & \ Ensemble \\
\ & \ B & \ C & \ D \\
\hline
\ $Z_{qq} \ (\mu=\nu)$ & \ 1.0982(3) \ & \ 1.1164(3) \ & \ 1.1564(1)\ \\
\ $Z_{qq} \ (\mu \neq \nu)$ & \ 1.1228(2) \ & \ 1.1481(3) \ & \ 1.1825(1)\ \\
\ $Z_{qq}^{\rm s} \ (\mu=\nu)$ & \ 1.0888(72) \ & \ 1.0987(99) \ & \ 1.1636(88)\ \\
\ $Z_{qq}^{\rm s} \ (\mu \neq \nu)$ & \ 1.0996(15) \ & \ 1.1323(12) \ & \ 1.1740(12)\ \\
\ $Z_{qg} \ (\mu \neq \nu)$ & \ -0.0106 \ & \ -0.0145 \ & \ -0.0188 \ \\
\ $Z_{gq} \ (\mu = \nu)$ & \ \ 0.0658 \ & \ \ 0.0410 \ & \ \ 0.0139 \ \\
\ $Z_{gq} \ (\mu \neq \nu)$ & \ \ 0.0772 \ & \ \ 0.0521 \ & \ \ 0.0246 \ \\
\ $Z_{gg}^{n_{\rm st} = 5} \ (\mu \neq \nu)$ & \ 0.739(8) \ & \ 0.727(21) \ & \ 0.792(35) \ \\
\ $Z_{gg}^{n_{\rm st} = 6} \ (\mu \neq \nu)$ & \ 0.744(7) \ & \ 0.738(19) \ & \ 0.799(32) \ \\
\ $Z_{gg}^{n_{\rm st} = 7} \ (\mu \neq \nu)$ & \ 0.755(7) \ & \ 0.753(18) \ & \ 0.812(30) \ \\
\ $Z_{gg}^{n_{\rm st} = 8} \ (\mu \neq \nu)$ & \ 0.767(7) \ & \ 0.769(17) \ & \ 0.826(28) \ \\
\ $Z_{gg}^{n_{\rm st} = 9} \ (\mu \neq \nu)$ & \ 0.778(6) \ & \ 0.784(16) \ & \ 0.839(26) \ \\
\ $Z_{gg}^{n_{\rm st} = 10} \ (\mu \neq \nu)$ & \ 0.790(6) \ & \ 0.797(16) \ & \ 0.850(25) \ \\
\hline
  \end{tabular}
  \label{tab:Zfactors}
\end{table}

\subsection{Lattice setup}
We provide additional information about the ETMC ensembles
\cite{Alexandrou:2018egz} in Tab.~\ref{tab:ensemble-statistics},
together with volume of statistics used in this work.

\begin{table}[h]
\caption{Statistics per ensemble used in this work. $n_{\mathrm{2 pt}}$ and $n_{\mathrm{3 pt}}$
give the number of stochastic samples produced per gauge configuration for two-point and (connected) three-point
functions, respectively. In both cases stats is the total statistics for the given ensemble.}
\begin{tabular}{ccccccc}
\hline
Label & Ensemble name  & $n_{\mathrm{conf}}$ &  $n_{\mathrm{2 pt}}$ & stats &  $n_{\mathrm{3 pt}}$ & stats \\
\hline
B & cB211.072.64 & 755 & 400  & 302,000 & 8 & 6040 \\
C & cC211.060.80 & 400 & 240  &  96,000 & 8 & 3200 \\ 
D & cD211.054.96 & 499 & 600  & 299,400 & 8 & 3992 \\ 
\hline
\end{tabular}
\label{tab:ensemble-statistics}
\end{table}

\begin{figure}[htpb]
  \centering
  \includegraphics[width=0.6\columnwidth]{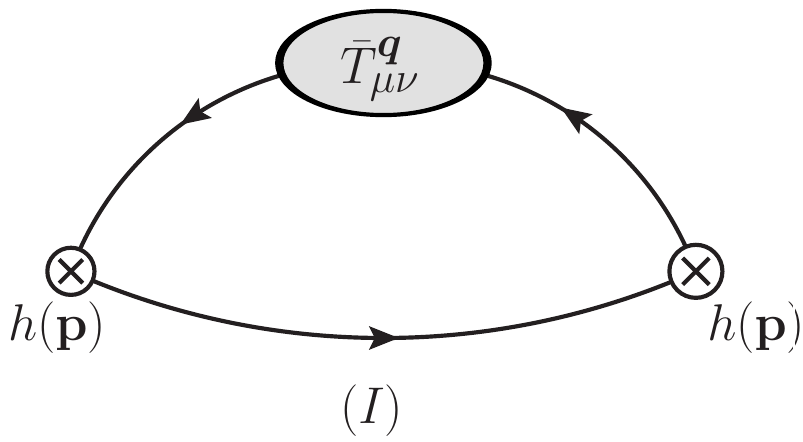} \\
  \bigskip
  \includegraphics[width=0.6\columnwidth]{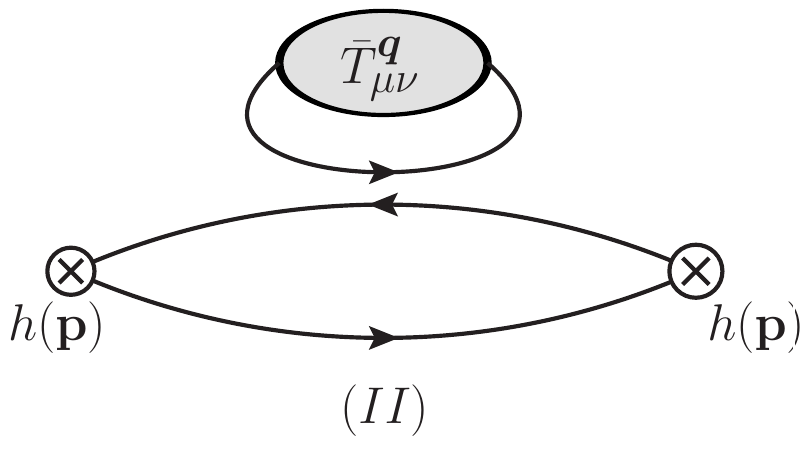} \\
  \bigskip
  \includegraphics[width=0.6\columnwidth]{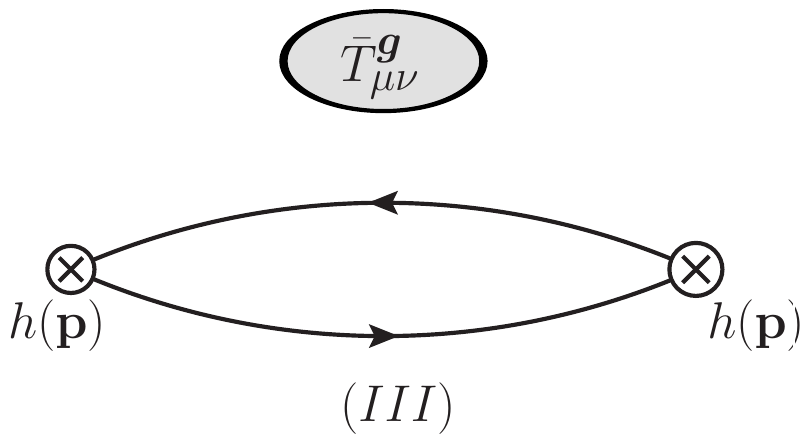}
  \caption{Quark-flow diagrams for three-point function with 
    quark and gluon energy momentum tensor operator insertion:
   $(I)$ connected and $(II)$ disconnected quark EMT insertion diagram,
  $(III)$ disconnected gluon EMT insertion diagram.
  }
  \label{fig:emt-3pt}
\end{figure}

The three different types of diagrams generated by the EMT insertions
$\bar T^{q}$ and $\bar T^{g}$ into the pion and kaon correlation
function are depicted in Fig.~\ref{fig:emt-3pt}.
We distinguish the (quark-)connected diagram $(I)$, and
the disconnected diagrams with a quark loop $(II)$ from inserting $\bar T^q$
and with the gluonic operator insertion $\bar T^{g}$ in $(III)$.

All quark-line (sub-)diagrams for three-point and two-point correlation functions
are produced with stochastic timeslice-to-all propagators with randomly chosen
source time\-slices. 
In case of the connected three-point function diagram $(I)$ we apply another sequential 
inversion through the hadron vertex at sink time.

To compute the diagrams involving strange or charm quark flow we use a mixed action for
the strange and charm quark: to avoid the technical complication of
strange and charm quark mixing inherent in the twisted mass sea quark action
we introduce additional valence quark doublets for both strange and charm flavor
in analogy to the light quark doublet with bare twisted masses of opposite sign within each doublet.
These bare twisted valence quark masses, $\mu_s$ and $\mu_c$, are tuned such that the 
$\Omega^-$ and
$\Lambda_c$ baryons take their physical values in the continuum limit,
leading to the  parameters given in Table~\ref{tab:mu-params}.
\begin{table}
\caption{The tuned strange $\mu_s$ and charm $\mu_c$ bare masses for the three ensembles}    \label{tab:mu-params}
    \begin{tabular}{lrrr}
    \hline
         &  B & C & D \\
         \hline
     $a\mu_s$    & $0.0186$ & $0.01615$ & $0.0136$\\
     $a\mu_c$    & $0.2490$ & $0.2060$ & $0.1660$ \\
      \hline
    \end{tabular}
\end{table}

Observables derived from the mixed action retain automatic $\mathcal{O}(a)$ improvement
with unitarity violations scaling equally at last at second order in the lattice spacing in the continuum limit. \\

Tables \ref{tab:ensemble-statistics} and \ref{tab:loopstats} list the number of stochastic samples for two- and three-point
functions per gauge configuration together with the total achieved statistics. \\

The quark loops for diagram $(II)$ with covariant derivative insertion are produced with
hierarchical probing and full dilution in spin and color degrees of freedom. For the light quark loops of ensembles B and C additional data sets were produced using
low-mode deflation on top of hierarchical probing. The parameters for the loop computation per gauge configuration are given in Table~\ref{tab:loopstats}. \\

\begin{table}
\caption{For each quark flavor we give in this order the number of deflated low modes $N_{\rm defl}$, the number of stochastic sources $N_{\rm srcs}$ and the probing distance $N_{col}$. We note that for the B and C ensembles, we have two sets of loops for the light quarks, which are averaged at the level of the analysis.}  

\label{tab:loopstats}
    \begin{tabular}{c||ccc|ccc|ccc}\hline
           \multirow{2}{*}{Ensemble} & \multicolumn{3}{c|}{Light loops} & \multicolumn{3}{c|}{Strange loops} & \multicolumn{3}{c}{Charm loops} \\
           & $N_{\rm defl}$ & $N_{\rm srcs}$ & $N_{\rm col}$ & $N_{\rm defl}$ & $N_{\rm srcs}$ & $N_{\rm col}$ & $N_{\rm defl}$ & $N_{\rm srcs}$ & $N_{\rm col}$ \\
         \hline
         B &  0 & 2 & 8 & 0 & 2 &8  & 0 & 12 & 4 \\
          & 200 &1 & 8 &  & & & & & \\
         \hline
         C & 0& 2 &8 & 0& 4 &8  & 0& 1 &8 \\
           &  450 & 1 & 8 & & & & & & \\
          \hline
         D & 0& 8 &8 & 0& 4 & 8 & 0& 1 &8 \\
         \hline         
    \end{tabular}
\end{table}

For completeness we give the connection between the ratios of two- and three-point functions
with EMT insertion $\bar T^{q},\,\bar T^{g}$ for tensor components used in
this work and the bare $\langle x \rangle$ values in Eq.~(\ref{eq:ratio-to-xbare}).
\begin{align}
  R_{44}^X(\tins, \tsnk;\, \mathbf{0}) & \overset{\tsnk,\tins \mathrm{~large}}{\longrightarrow}
  -\frac{3}{4}\, \langle x \rangle_{X} \,m_h \,,
  \nonumber \\
  R_{4k}^X(\tins, \tsnk;\, \pvec) & \overset{\tsnk,\tins \mathrm{~large}}{\longrightarrow}
  \langle x \rangle_{X} \,p_k \,.
  \label{eq:ratio-to-xbare}
\end{align}
In both cases the relation holds up to excited state contamination and up to the finite-time wrapping correction factor $[1+\exp(-E_h(T - 2\,\tsnk))]^{-1}$
as given in Eq.~(\ref{eq:RX}).

\end{document}